\documentclass[]{pasj01_revised}
\begin{document}

\title{Resolving 4-D Nature of Magnetism with Depolarization and Faraday Tomography: Japanese SKA Cosmic Magnetism Science}
\author{
Takuya \textsc{Akahori}\altaffilmark{1}\thanks{corresponding author; akahori@sci.kagoshima-u.ac.jp},
Yutaka \textsc{Fujita}\altaffilmark{2},
Kiyotomo \textsc{Ichiki}\altaffilmark{3},
Shinsuke \textsc{Ideguchi}\altaffilmark{4},
Takahiro \textsc{Kudoh}\altaffilmark{5},
Yuki \textsc{Kudoh}\altaffilmark{6},
Mami \textsc{Machida}\altaffilmark{7},
Hiroyuki \textsc{Nakanishi}\altaffilmark{1},
Hiroshi \textsc{Ohno}\altaffilmark{8},
Takeaki \textsc{Ozawa}\altaffilmark{1},
Keitaro \textsc{Takahashi}\altaffilmark{9},
Motokazu \textsc{Takizawa}\altaffilmark{10}, 
on behalf of the SKA-JP Magnetism SWG.
}

\altaffiltext{1}{Graduate School of Science and Engineering, Kagoshima University, Kagoshima 890-0065}
\altaffiltext{2}{Graduate School of Science, Osaka University, Osaka 560-0043}
\altaffiltext{3}{Kobayashi-Maskawa Institute, Nagoya University, Aichi 464-8602}
\altaffiltext{4}{Department of Physics, UNIST, Ulsan 44919, Korea}
\altaffiltext{5}{Faculy of Education, Nagasaki University, Nagasaki 852-8521}
\altaffiltext{6}{Faculty of Sciences, Chiba University, Chiba-shi 263-8522}
\altaffiltext{7}{Faculty of Sciences, Kyushu University, Fukuoka 812-8581}
\altaffiltext{8}{Tohoku Bunkyo College, Yamagata 990-2316}
\altaffiltext{9}{Department of Physics, Kumamoto University, Kumamoto 860-8555}
\altaffiltext{10}{Department of Physics, Yamagata University, Yamagata 990-8560}

\KeyWords{magnetic fields --- polarization}

\maketitle

%%%%%%%%%%%%%%%%%%%%%%%%%%%%%%%%%%%%%%%%%%%%%%%
%%%%%%%%%%%%%%%%%%%%%%%%%%%%%%%%%%%%%%%%%%%%%%%
%%%%%%%%%%%%%%%%%%%%%%%%%%%%%%%%%%%%%%%%%%%%%%%
\begin{abstract}

Magnetic fields play essential roles in various astronomical objects. Radio astronomy has revealed that magnetic fields are ubiquitous in our Universe. However, the real origin and evolution of magnetic fields is poorly proven. In order to advance our knowledge of cosmic magnetism in coming decades, the Square Kilometre Array (SKA) should have supreme sensitivity than ever before, which provides numerous observation points in the cosmic space. Furthermore, the SKA should be designed to facilitate wideband polarimetry so as to allow us to examine sightline structures of magnetic fields by means of depolarization and Faraday Tomography. The SKA will be able to drive cosmic magnetism of the interstellar medium, the Milky Way, galaxies, AGN, galaxy clusters, and potentially the cosmic web which may preserve information of the primeval Universe. The Japan SKA Consortium (SKA-JP) Magnetism Science Working Group (SWG) proposes the project {\it "Resolving 4-D Nature of Magnetism with Depolarization and Faraday Tomography"}, which contains ten scientific use cases.

\end{abstract}

\begin{figure}
\vspace{-210mm}%%%'±'±'̐"'l'͑̍قð•Ï'¦'È'¢'l'ðŽè'T'è'Å'T'·
\includegraphics[width=25mm]{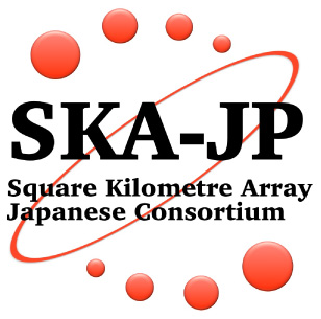}
\vspace{177mm}%%%'±'±'̐"'l'͑̍قð•Ï'¦'È'¢'l'ðŽè'T'è'Å'T'·
\end{figure}

%%%%%%%%%%%%%%%%%%%%%%%%%%%%%%%%%%%%%%%%%%%%%%%
%%%%%%%%%%%%%%%%%%%%%%%%%%%%%%%%%%%%%%%%%%%%%%%
%%%%%%%%%%%%%%%%%%%%%%%%%%%%%%%%%%%%%%%%%%%%%%%
\section{Introduction}
\label{mag.s1}

(Electro)-Magnetism is one of the four fundamental forces in the Universe. Magnetic fields induce fundamental astrophysical processes such as particle acceleration, nonthermal radiation, and polarization, and impact on activities of astronomical objects through field tension, reconnection, instability, and turbulence. It is interesting that such rich, diverse nature of magnetic fields is often explained by common theories of magnetism, though various effects caused by magnetic fields make them difficult to understand themselves. Understanding local magnetic fields will assist studies of the epoch of reionization, the cosmic microwave background polarization, and the ultra high energy cosmic rays (CRs), while deepest magnetic fields may preserve information of the early Universe. 

Centimeter and meter wavelength radio astronomy is a pioneer and leader of exploring magnetic fields in the Universe. It has discovered magnetic fields in and around most of astronomical objects, and has provided a lot of clues for understanding the nature of magnetic fields. However, the real origin and evolution of magnetic fields is poorly proven, because it is difficult with a lack of sensitivity and polarization capability to go back in time by exploring high redshift Universe. Even magnetic fields in the Milky Way is not well-understood, because a wide-field survey is essential but is also expensive. These challenges suggest an observatory having supreme sensitivity and high polarization capability on the scale of the SKA. 

We, the SKA-JP Magnetism SWG, have developed our science objectives since 2010. We conclude that comprehensive study from the Sun to the early Universe is important to understand magnetism's diversity and universality. Therefore, instead of making a short list of science objectives, we focus on SKA's capability of wideband polarimetry which enable us to utilize some new observational techniques introduced in the following sections. We propose the project {\it ''Resolving 4-D Nature of Magnetism with Depolarization and Faraday Tomography"}, in which we demonstrate ten scientific use cases based on careful considerations on depolarization and Faraday tomography.

The purpose of this document is to introduce Japanese scientific interests in the SKA project and to report results of our investigation. It is still in progress, so that the document may not fully cover previous works related to the SKA. We wish that the document becomes an interface for future communications, collaborations, and synergies with worldwide communities. In the rest of this section, we introduce the methods to study magnetic fields with the SKA, except the Zeeman effect which is introduced in a separate paper. We overview the international magnetism sciences in Section 2 and describe Japanese magnetism sciences in Section 3. We summarize in Section 4.

%%%%%%%%%%%%%%%%%%%%%%%%%%%%%%%%%%%%%%%%%%%%%%%
%%%%%%%%%%%%%%%%%%%%%%%%%%%%%%%%%%%%%%%%%%%%%%%
\subsection{Synchrotron Radiation and Faraday Rotation}
\label{mag.s1.ss1}

Below, so as to make discussion easy, we figure out the minimum basis and physical parameters of synchrotron radiation and Faraday rotation, which are well-established tools to study cosmic magnetism. See the textbooks for details (e.g., \cite{1979rpa..book.....R,1996tra..book.....R}).

Synchrotron radiation is the bremsstrahlung emitted from a relativistic particle gyrating around magnetic fields. Suppose the isotropic distribution of relativistic electrons, we consider the following energy distribution of the electrons, $N(\gamma)d\gamma=C(r)\gamma^{-p}d\gamma$, where $\gamma$ is the Lorentz factor, $C(r)$ is the number density of the electrons located at the position $r$, and $p$ is the spectral index. The synchrotron emissivity can be written as 
\begin{equation}
\epsilon (r) \propto C(r)B_\perp(r)^{(1+p)/2}\omega^{(1-p)/2}
\end{equation}
where $B_\perp$ is the strength of magnetic fields perpendicular to the line-of-sight (LOS) and $\omega = 2 \pi \nu$ is the angular frequency. The integral of the emissivity gives the Stokes parameters $I$, $Q$, $U$, $V$, and we define the linear polarization, $P=Q+iU$, the fractional polarization, $\Pi=(Q^2 + U^2 + V^2)^{1/2}/I$, the polarization angle, $\chi=(1/2)\arctan(U/Q)$, and the electric-vector angle $\psi=2\chi$, where $\chi$ ranges $-\pi/4 \le \chi \le \pi/4$. Therefore, once we observe synchrotron radiation and assume $C(r)$, we can estimate the integral of $B_\perp$.

When polarized emission passes through magneto-ionic media, its polarization angle rotates as $\chi=\chi_0+RM\lambda^2$ (Faraday rotation). Here $\chi_0$ is the initial polarization angle and $\lambda$ is the wavelength. RM is the rotation measure, 
\begin{equation}
RM~{\rm (rad~m^{-2})} \approx 811.9 \int 
\left(\frac{n_{\rm e}}{\rm cm^{-2}}\right)
\left(\frac{B_{||}}{\rm \mu G}\right)
\left(\frac{dr}{\rm kpc}\right),
\end{equation}
where $n_{\rm e}$ is the density of free electrons, $B_{||}$ is the strength of magnetic fields parallel to the LOS, and RM is positive if the magnetic field points toward the observer. Therefore, once we observe RM and assume $n_{\rm e}$, we can estimate the integral of $B_{||}$. Faraday rotation suggests that RM can be derived from at least two polarization angles measured at different wavelengths. In addition, with RM, we can perform inverse Faraday rotation and can guess the intrinsic polarization angle, which is perpendicular to the field direction if the polarization is synchrotron.

%%%%%%%%%%%%%%%%%%%%%%%%%%%%%%%%%%%%%%%%%%%%%%%
%%%%%%%%%%%%%%%%%%%%%%%%%%%%%%%%%%%%%%%%%%%%%%%
\subsection{Depolarization and Faraday Tomography}
\label{mag.s1.ss2}

Depolarization and Faraday tomography are thought to have the capability to resolve three dimensional structure of magnetic fields. But their applications to real observations have been limited and they have been under development since 60's (e.g., \cite{1966MNRAS.133...67B}), because they require wideband polarimetric data which are expensive to obtain. The SKA should facilitate wideband polarimetric observation and break through the situation in order to advance the study of cosmic magnetism.

\begin{figure}[tbp]
\begin{center}
\FigureFile(65mm,65mm){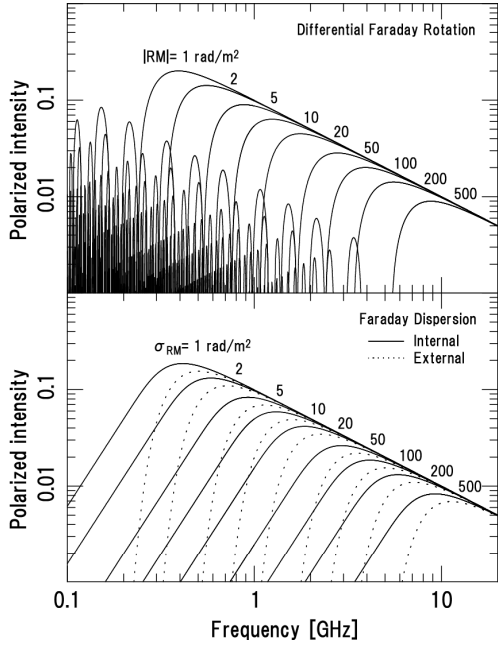}
\end{center}
\caption{
Depolarization of differential Faraday rotation (top) and dispersion (bottom). See \citet{2011MNRAS.418.2336A} for the original figures and details.
}
\label{f01}
\end{figure}

Depolarization is the phenomenon that we observe weaker polarization than that at the origin. It depends on properties of matters along the path of the polarization. \citet{1966MNRAS.133...67B} formulated depolarization in several simple situations. For example, in the case that there is a matter with uniform electron density and uniform magnetic fields, the fractional polarization $\Pi_l$ of the matter for a specific sightline can be written as 
\begin{equation}
\Pi_l(\lambda^2) = \Pi_{l0}(\lambda^2)
\frac{\sin \left( RM \lambda^2 \right) }{RM \lambda^2}
\exp{2i \left(  \chi_0 + \frac{1}{2}RM \lambda^2 \right) },
\end{equation}
where $\Pi_{l0}$ is the intrinsic fractional polarization. This case is called as differential Faraday rotation depolarization. 

Another example is beam depolarization. We suppose that RM distribution inside a beam area follows the Gaussian with the standard deviation of $\sigma_{\rm RM}$. In the case that this RM source itself is an emitter of polarization, $\Pi_l$ can be written as 
\begin{equation}
\Pi_l(\lambda^2) = \Pi_{l0}(\lambda^2)
\frac{1-\exp \left(-2\sigma_{\rm RM}^2 \lambda^4\right)}{2\sigma_{\rm RM}^2 \lambda^4},
\end{equation}
and is called as internal Faraday dispersion depolarization. Otherwise, in the case that this RM source is a foreground of background polarization, $\Pi_l$ can be written as 
\begin{equation}
\Pi_l(\lambda^2) = \Pi_{l0}(\lambda^2)\exp \left(-2\sigma_{\rm RM}^2 \lambda^4\right),
\end{equation}
and is called as external Faraday dispersion depolarization. 

Figure \ref{f01} shows the Burn laws. As shown, the degree of depolarization depends on frequency. Wide frequency coverage is hence essential to capture the feature of depolarization. The fact that depolarization is weaker at higher frequencies implies that the observed fractional polarization can tend to be larger for higher redshift sources, because depolarization happens at their rest-frame (higher) frequencies.

Faraday Tomography is a decomposition technique proposed by \citet{1966MNRAS.133...67B}. The polarized intensity can be written as 
\begin{equation}
P(\lambda^2)=\int^\infty_{0} \varepsilon(r)e^{2i\chi(r,\lambda^2)}dr
=\int^\infty_{-\infty}F(\phi)e^{2i\phi\lambda^2}d\phi,
\end{equation}
where $\chi(r,\lambda^2)=\chi_0(r)+\phi(r)\lambda^2$ is the polarization angle at the observer, $\chi_0(r)$ is the polarization angle at the origin $r$, and $\phi(r)$ is the Faraday depth in rad~m$^{-2}$. $F(\phi) \equiv \varepsilon(\phi)e^{2i\chi_0(\phi)}$ is the Faraday dispersion function (FDF) or the Faraday spectrum, which represents cross sections of the polarized intensity along the LOS (though $\phi(r)$ and $r$ are not in one to one correspondence). We changed the integration variable from $r$ to $\phi(r)$, resulting in the form of Fourier transformation with conjugate variables $\phi$ and $\lambda^2$. Therefore, the inverse Fourier transformation can derive $F(\phi)$ from the observable $P(\lambda^2)$. 

\begin{figure}[tbp]
\begin{center}
\FigureFile(80mm,80mm){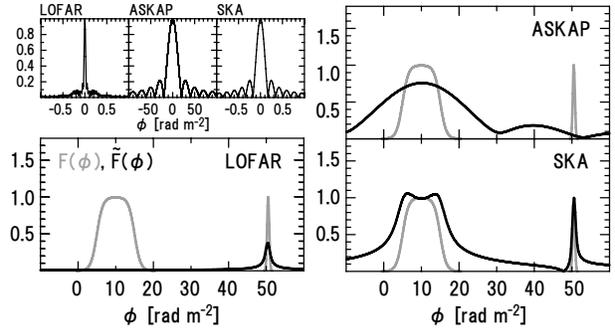}
\end{center}
\caption{
The rotation measure spread function (RMSF, top left), and reconstructed Faraday spectra \citep{2014PASJ...66...65A}. LOFAR and ASKAP frequency coverages are sensitive to only thin and thick Faraday structure, respectively, while SKA's seamless, broad bandwidth can provide sensitivities for RM structures of $O(1-1000)$ rad~m$^{-2}$. 
}
\label{f02}
\end{figure}

Practically, using a window function $W(\lambda^2)$ for observable wavelengths, the reconstructed FDF, $\tilde F(\phi)$, can be written as 
\begin{equation}
\tilde F(\phi)=\frac{1}{2\pi}\int_{-\infty}^\infty W(\lambda^2)P(\lambda^2)e^{-2i\phi\lambda^2}d\lambda^2.
\end{equation}
Using the convolution theorem, this equation becomes $\tilde F(\phi)=K^{-1}R(\phi)\ast F(\phi)$ and $R(\phi)=K\int^\infty_{-\infty} W(\lambda^2)e^{-2i\phi\lambda^2}d\lambda^2$, where $K$ is the normalization and $R(\phi)$ is the rotation measure spread function (RMSF). If $R(\phi)/K$ is the delta function $\delta(\phi)$, the reconstruction is perfect. But in reality it is impossible to fill out all $\lambda^2$ space with observed data. Figure~\ref{f02} shows some examples of RMSF.

The full width at half maximum (FWHM) of the RMSF corresponds to the resolution in the Faraday depth space. For example, we consider the case that the window function is $W(\lambda^2)=1$ at observed wavelengths ($\lambda^2_{\rm min} \leq \lambda^2\leq \lambda^2_{\rm max}$) and otherwise $W(\lambda^2)=0$. The FWHM of the RMSF is
\begin{equation}
{\rm FWHM~(rad~m^{-2})} = \frac{2\sqrt3}{\Delta\lambda^2({\rm m^2})}.
\end{equation}
This indicates that the resolution is determined by the $\lambda^2$-space coverage, $\Delta\lambda^2 = \lambda^2_{\rm max} - \lambda^2_{\rm min}$. Particularly, going down to a longer wavelength expands the $\lambda^2$ coverage effectively, though it tends to suffer from depolarization more seriously. Depolarization can be significant at the wavelength where the polarization angle rotates by $\pi$. Hence the maximum observable Faraday depth width of a feature which is extended in the Faraday depth space is 
\begin{equation}
L_{\rm \phi, max}~{\rm~(rad~m^{-2})} \approx \frac{\pi}{\lambda^2_{\rm min}({\rm m^2})}
\end{equation}
In order to make Faraday tomography feasible, the above two frequency conditions are at least needed to be improved.

%%%%%%%%%%%%%%%%%%%%%%%%%%%%%%%%%%%%%%%%%%%%%%%
%%%%%%%%%%%%%%%%%%%%%%%%%%%%%%%%%%%%%%%%%%%%%%%
%%%%%%%%%%%%%%%%%%%%%%%%%%%%%%%%%%%%%%%%%%%%%%%
\section{SKA Magnetism Science}
\label{mag.s2}

\begin{table*}
\tbl{Cosmic magnetism priority science objectives. The re-baseline of the SKA phase 1 (SKA1) is not incorporated.}{%
\begin{tabular}{p{23em}llllll}
\hline
Title & Field of view (FoV) & Band & Sensitivity & Resolution & Pol. Purity\\
\hline
The resolved all-sky characterization of the interstellar and intergalactic magnetic fields & all-sky (1.5 yrs) & 1--1.5 GHz & 2~$\mu$Jy & $2''$ & 0.1~\% \\
Determine origin, maintenance and amplification of magnetic fields at high redshifts - I. & 10 deg$^2$ & 1--2 GHz & 0.1~$\mu$Jy & $0.''5$--$1.''0$ & 0.1~\% \\
Detection of polarized emission in Cosmic Web filaments & Targets & 600~MHz & 0.1~$\mu$Jy & $0.''5$--$1.''0$ & 0.1~\% \\
Determine origin, maintenance and amplification of magnetic fields at high redshifts - II. & 3 deg$^2$ & 2--3 GHz & 0.1~$\mu$Jy & $0.''5$--$1.''0$ & 0.1~\% \\
Intrinsic properties of polarized sources & 100 hrs/target & 3--5 GHz & 0.2~$\mu$Jy & $5.''0$ & 0.1~\% \\
\hline
\end{tabular}}\label{t01}
\end{table*}

\begin{table*}
\tbl{Cosmic magnetism notional Key Science Projects (KSP) made at the KSP meeting 2015 at Stockholm (E: essential, D: desirable). There are two KSPs: 1. the origin and evolution of magnetic fields in large-scale structure (KSP1, the top three rows), and 2. the origin and evolution of magnetic fields in galaxies (KSP2, the bottom eight rows).}{%
\begin{tabular}{p{25em}llllll}
\hline
& All-sky& & & & & \\
Title & Band-2 & LOW & B1 & B2 & B5 & Area\\
\hline
The magnetic field in clusters and filaments &
E & E & E & - & - & Band-1 60 deg$^2$, low all-sky \\
Probing the nature of Dark Matter &
E & - & E & E & E & 10 fields \\
The magnetic cosmic web &
D & - & E & E & - & 100 deg$^2$\\
\hline
LOS probes of evolution of cosmic magnetism &
E & - & - & - & - & all-sky \\
Evolution of the magnetic fields in galaxy disks &
- & - & - & E & - & 10 deg$^2$ \\
Broad-band polarimetry as a probe of AGN &
E & E & E & E & - & all-sky \\
Magnetic fields in AGN at all redshifts and luminosities &
E & - & - & - & VLBI & VLBI follow up \\
Emergence of magnetic fields in the Universe &
D & E & E & - & - & all-sky \\
Magnetic fields in nearby galaxies &
D & - & - & E & E & 25--30 galaxies \\
Magnetic fields in the heart of the Milky Way &
D & - & E & E & E & 4000 deg$^2$ \\
Multi-scale magnetism in the Milky Way &
E & - & E & - & - & 30000 deg$^2$ \\
\hline
\end{tabular}}\label{t02}
\end{table*}

In this section, we briefly introduce major magnetism science partly developed by the international SKA Cosmic Magnetism SWG, and partly presented in the international SKA Science Book (2015). The SWG are also developing notional survey projects, which are summarized in Tables \ref{t01} and \ref{t02}. A polarization sky survey with SKA1-MID Band-2 is the highest priority science objective, which maximizes scientific outcomes including local clouds to high redshift objects.

\begin{figure}[tbp]
\begin{center}
\FigureFile(75mm,75mm){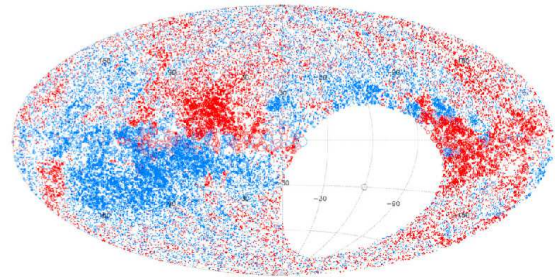}
\end{center}
\caption{
The longitude-latitude map of RMs obtained from the NRAO VLA Sky Survey (NVSS, \cite{2009ApJ...702.1230T}). The size and color of each circle shows the amplitude of RM and its sign (red: positive, blue: negative), respectively, of each extragalactic source.
}
\label{f03}
\end{figure}

Figure~\ref{f03} shows the largest (37,543) RM catalog of extragalactic polarized sources \citep{2009ApJ...702.1230T}. All-sky polarization surveys can provide such ``RM grid'' maps, which are useful for studying spatial structures of magnetic fields. Increasing the number of sources, i.e. denser RM grids, can reveal smaller spatial structures, and can also provide more observation points in redshift space, allowing us to study the cosmic evolution of magnetic fields.

\begin{figure}[tbp]
\begin{center}
\FigureFile(75mm,75mm){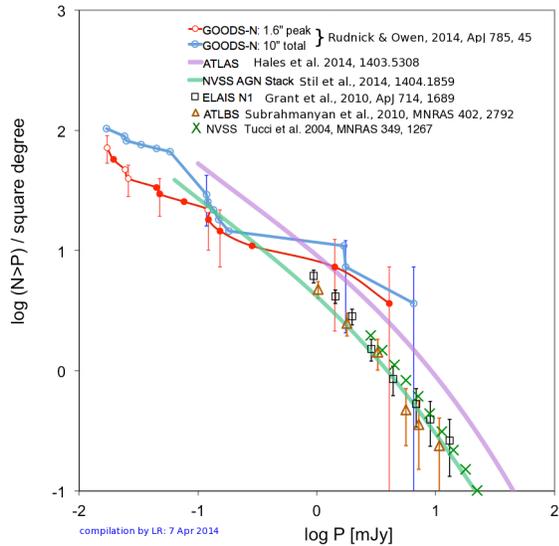}
\end{center}
\caption{
Population of polarized sources observed at 1.4 GHz (\cite{2014ApJ...785...45R, 2015aska.confE.113T}). The red and blue, purple, and green lines are based on the deep observations of the $\sim 0.2$ deg$^2$ VLA GOODS-N field \citep{2014ApJ...785...45R}, the $\sim 6$ deg$^2$ ATCA ATLAS field \citep{2014MNRAS.440.3113H}, and the all-sky VLA NVSS survey \citep{2014ApJ...787...99S}, respectively.
}
\label{f04}
\end{figure}

Table~\ref{t03} lists the expected density of RM grids to be obtained with the SKA1. These estimations are basically derived from Figure~\ref{f04}, which summarizes plots of observed source populations (mostly FR galaxies and quasars) at 1.4 GHz. Because the measured populations have a large scatter, there is still a factor of four uncertainty in the estimation of RM grids  \citep{2014skao.rept.....G}; the population below 10 $\mu$Jy is not well-known. Faint components such as star-forming galaxies, ULIRGs, merging galaxies, and quiet spiral galaxies, would be more important in the low luminosity regime \citep{2015aska.confE.113T}. Deep pilot observations (Fig. \ref{f04}) are useful to foresee the density of RM grids in the SKA era. Studying the population itself is an interesting theme to understand the population of galaxies and AGN.

\begin{table}
\tbl{The expected density of RM grids at 1.4 GHz}{%
\begin{tabular}{rrl}
\hline
Pol. Intensity & Counts$^\dagger$ & Reference \\
\hline
4 $\mu$Jy & $\sim$230--450/deg$^2$ & \citet{2015aska.confE..92J}\\
0.75 $\mu$Jy & $\sim$5000/deg$^2$ & \citet{2015aska.confE.113T}\\
\hline
\end{tabular}}\label{t03}
\begin{tabnote}
$^\dagger$The number of extragalactic polarized sources.
\end{tabnote}
\end{table}

%%%%%%%%%%%%%%%%%%%%%%%%%%%%%%%%%%%%%%%%%%%%%%%
%%%%%%%%%%%%%%%%%%%%%%%%%%%%%%%%%%%%%%%%%%%%%%%
\subsection{Interstellar Medium}
\label{mag.s2.ss1}

SKA's dense RM grids will allow us to study magnetic-field structures of various discrete sources in detail. Furthermore, we will be able to study the structures along the LOS through Faraday tomography with broadband polarimetry. Below, we briefly summarize the current status of research in magnetic fields of various discrete sources in the Milky Way. See e.g., \citet{2015aska.confE..96H} and references therein.

\begin{itemize}

\item{\bf Young stellar objects (YSOs):} Jets and outflows of YSOs have been studied through observations of molecular emission lines. MHD processes are widely accepted as the launching mechanisms for the jets and the outflows. Measurement of linearly polarized synchrotron emission makes it possible to study the magnetic-field structure of YSOs.

\item{\bf Supernova remnants (SNRs):} A SNR expands into the interstellar medium (ISM) with kinetic energy of $\sim 10^{51}$ ergs. The kinetic energy can be converted to the energies of magnetic fields, high-energy particles, and turbulence. Actually, observations suggest that magnetic fields in SNRs should be amplified to mG through interaction between shock waves and the ISM. Shock waves and magnetic fields are important for acceleration of the Galactic CRs. SNRs are thought to be the main energy source which drives the interstellar turbulence.

\item{\bf HII regions:} The strength of magnetic fields in HII regions has been estimated to be a few to a few tens of $\mu$G from observations of Faraday rotation of background radio sources and background Galactic synchrotron emission \citep{2007A&A...463..993S, 2010A&A...515A..64G, 2011ApJ...736...83H}. It is important to understand the dynamical role of the magnetic fields in HII regions over a large density range. We need to measure magnetic fields in HII regions using dense RM grids.

\item{\bf Planetary nebulae:} Old planetary nebulae are good probes of the ISM magnetic fields. The ISM magnetic field around the nebula shell is compressed and deflected. Since background radio emission is Faraday rotated with this field, we can observe systematic variation of the polarization angle across the shell. It would be possible to study the ISM magnetic fields around the planetary nebulae through measurements of polarization with high angular resolution.

\item{\bf Pulsar wind nebulae (PWNe):} The pulsar generates a relativistic magnetohydrodynamic (MHD) wind. Such pulsar wind terminates at a strong shock. Relativistic particles in the downstream of the shock emit non-thermal radiations,  which are identified as PWNe. It would be possible to understand the role of magnetic fields through observations of polarized emission from PWNe and RMs of background sources.

\item{\bf Faraday screens:} Faraday screens are unidentified radio sources. They are visible in polarized intensity and RM, but are faint in total intensity. Some of them may be associated with PWNe and old HII regions.

\item{\bf Molecular clouds:} The role of magnetic fields in star formation has been debated for decades. Strong- and weak-field models, in which the efficiency of ambipolar diffusion is different, predict different star-formation scenarios. While it is difficult to measure the detailed field strength and geometry of dense molecular clouds, the SKA could detect in-situ synchrotron radiation of $\sim$ mJy at 1 GHz emitted by Galactic CRs penetrating collapsing dense molecular clouds.

\end{itemize}

{\it Related SKA Science Book 2015 Chapters; \citet{2015aska.confE..41H}, \citet{2015aska.confE..96H}, \citet{2015aska.confE.102D}, \citet{2015aska.confE.110R}.}

%%%%%%%%%%%%%%%%%%%%%%%%%%%%%%%%%%%%%%%%%%%%%%%
%%%%%%%%%%%%%%%%%%%%%%%%%%%%%%%%%%%%%%%%%%%%%%%
\subsection{The Milky Way}
\label{mag.s2.ss2}

The Milky Way has large-scale magnetic fields along the spiral arms. The typical strength of the magnetic fields is about 3 $\mu$G (e.g., \cite{2006ApJ...642..868H}). Recent observational studies concluded that it has an axisymmetric spiral structure with at least one reversal in the Milky Way. The dynamo theory indeed favored an axisymmetric structure of even parity with respect to the galactic mid-plane, i.e., the horizontal field component has the same direction on either side of the mid-plane, while the vertical component changes sign across the mid-plane \citep{1988Natur.336..341R}. The toroidal field component is dominated in the axisymmetric structure together with a small radial component.

As for the Milky Way halo, some observations suggested that the magnetic-field structure is an dipolar-type (odd parity) \citep{1997A&A...322...98H}, spiral, or X-shaped, though it is still difficult to characterize the global halo structure. The dynamo theory favored axisymmetric mode with odd parity in the spherical objects such as a galactic halo. The X-shaped structure may be produced by a large-scale Galactic wind.

The turbulent magnetic field would exist because the turbulent gas exists in the ISM, although the power spectrum of the magnetic field is difficult to observe. The energy injection scale of the interstellar turbulence is suggested to be the order of 1~pc at spiral arms and the order of 100~pc at interarm regions.

The magnetic field strength at the Galactic center shows several values from 10~$\mu$G to 1~mG, depending on the scales. It is reported that the poloidal magnetic field is dominated around 50~pc to 300~pc near the dynamical center of the Milky Way.

Resolving magnetic fields in the Milky Way is essential to elucidate their origin and maintenance mechanism by comparing to theoretical models. Understanding structure of the magnetic fields is also crucial for the studies of CMB polarization, redshifted HI emission at the epoch of reionization (EoR), and deflection of ultra-high energy CRs, and so on.

{\it Related SKA Science Book 2015 Chapters; \citet{2015aska.confE..41H}, \citet{2015aska.confE..92J}, \citet{2015aska.confE..96H}.}

%%%%%%%%%%%%%%%%%%%%%%%%%%%%%%%%%%%%%%%%%%%%%%%
%%%%%%%%%%%%%%%%%%%%%%%%%%%%%%%%%%%%%%%%%%%%%%%
\subsection{Nearby Galaxies}
\label{mag.s2.ss3}

Similar to the Milky Way, galactic magnetic fields can be described as the combination of large-scale, regular fields and small-scale, turbulent fields. Observed depolarization implies that the coherence length of the turbulent magnetic fields is $\sim 50$~pc \citep{2007A&A...470..539B}. The regular magnetic fields are along with spiral structure and ordered in the regions where strong shear occurs, the ISM is compressed, or turbulence is small. X-shaped halo magnetic fields are seen in some galaxies \citep{2014arXiv1401.1317K,2011A&A...531A.127S}. They may link with the weak poloidal magnetic fields predicted by the $\alpha$-$\Omega$ dynamo theory, as well as galactic winds related to star formation.

Galactic magnetic fields are thought to be determined with parameters characterizing galactic large-scale structures, such as (i) the rotation curve related with shear motion of the ISM, (ii) the star formation rate related with the $\alpha$ effect, (iii) the density and turbulent velocity of the ISM indicating the upper limit of magnetic energy, and (iv) the disk size indicating magnetic order. However, there is no observational evidence supporting the correlation between large-scale magnetic fields and the star-formation rate. There is also no observational evidence supporting the prediction that turbulence dynamo worked significantly in the stage of galaxy formation. A clue is that reversal in magnetic fields due likely to turbulence dynamo is found in the Milky Way \citep{2011ApJ...728...97V}. 

Magnetic fields in elliptical and dwarf galaxies are highly unknown. It is true that there is differential rotation in both late- and early-type galaxies. The differential rotation can cause turbulence dynamo and the magneto-rotational and the Parker instabilities, which can produce both small- and large-scale magnetic fields. Therefore, both magnetic fields should exist in elliptical galaxies. However, observational study is difficult for elliptical galaxies because synchrotron emission is weak. A more fundamental problem would be that large-scale magnetic fields are observed in dwarf galaxies LMC, SMC, and IC10. It may be difficult to explain them by galactic dynamo since the dynamo does not work so much in such dwarf systems.

In order to resolve the problems described above and to understand the origin of galactic magnetic fields, it is essential to investigate many nearby galaxies and to examine diversity of magnetic fields and relations between magnetic fields and star-formation or galaxy morphology. The cosmological evolution of magnetic fields can be partly studied within a given redshift range. The international SWG proposed the following high resolution observation; (i) frequency range: 2.8--5.18~GHz with SKA1-MID Band-4, (ii) resolution: $5''$, (iii) sensitivity: 0.2~$\mu$Jy/beam (12 hrs) in rms. Targets are galaxies within 20~Mpc and Dec $<15^\circ$ listed in Table \ref{t04}.

\begin{table}
\tbl{Target Nearby Galaxies}{%
\begin{tabular}{ll}
\hline
Face-on & M33, NGC300, 628, 1566, 1808, 2997, Circinus A \\
Edge-on & M104, NGC55, 253, 3628, 4666, 4945 \\
Bar & M83, NGC107, 1313, 1365, 1502, 1672, 2442 \\
Irregular & LMC, SMC \\
Dwarf & NGC1140, 1705, 5253, IC4662 \\
Elliptical$^\dagger$ & NGC1404, 4697\\
\hline
\end{tabular}}\label{t04}
\begin{tabnote}
$^\dagger$Quiescent elliptical.
\end{tabnote}
\end{table}

{\it Related SKA Science Book 2015 Chapters; \citet{2015aska.confE..92J}, \citet{2015aska.confE..94B}, \citet{2015aska.confE.106H}.}

%%%%%%%%%%%%%%%%%%%%%%%%%%%%%%%%%%%%%%%%%%%%%%%
%%%%%%%%%%%%%%%%%%%%%%%%%%%%%%%%%%%%%%%%%%%%%%%
\subsection{Distant Galaxies}
\label{mag.s2.ss4}

As already described in the previous sections, dynamo is one of the key mechanisms for the generation and maintenance of galactic magnetic fields. Numerical simulations suggest that, while small-scale magnetic fields is established in hundreds million years through turbulence dynamo \citep{2012PhRvE..85b6303S}, large-scale magnetic fields grows up by the $\alpha$-$\Omega$ dynamo in the timescale comparable to the disk rotation \citep{2005PhR...417....1B}; for instance, kpc-scale magnetic fields at $z=3$ could evolve to galactic-scale fields by $z=0.5$ \citep{2009A&A...494...21A}. The evolution could be affected by the star-formation, feedback, and galaxy merger, although these factors are still hard to fully incorporate into cosmological MHD simulations of galaxy formation. 

Observing distant galaxies allows us to go back in time and to examine the evolution of galactic magnetic fields. However, it is difficult to observe faint galactic synchrotron emission from distant galaxies. In order to overcome the difficulty, three approaches are proposed in the SKA science book 2015.

The first approach is to carry out deep observations. We request (i) wideband to resolve a degeneracy between Faraday rotation and depolarization caused by multiple components along the LOS, (ii) high sensitivity and wide FoV to overlap with optical surveys, and (iii) high angular resolution to cut out an area of polarization corresponding to the position of the optical spectrum. With 100~hours per pointing for SKA1-MID Band-2, we will achieve the sensitivity 75~nJy in rms, which is one order of magnitude better than ever before. With the sensitivity, we can study the cosmological evolution of disk galaxies beyond the redshift $z>2.5$ and AGN and star-forming galaxies beyond $z>7$ \citep{2015aska.confE.113T}. The survey can combine with a commensal HI absorption survey and future optical and infrared surveys (4MOST, WFIRST, LSST, Euclid), allowing us to understand galactic gas, turbulence, magnetic fields, and their spatial distributions for a wide range of redshift.

The second approach is to perform stacking analysis, in which polarization signals of faint sources are cumulated based on their coordinates given by other surveys (Stokes I, optical, etc). The cumulation permits us to extract statistical information of sources whose flux are below the detection limit. For example, \citet{2014ApJ...787...99S} revealed that fainter sources below the detection limit of the NVSS $S_{1.4}\sim 80$~mJy tend to have higher fractional polarization. It may be because depolarization is minor at rest-frame frequencies of distant galaxies \citep{2002A&A...396..463M}. With stacking analysis, we could compare magnetic fields in dwarf galaxies, star-forming galaxies, and the Milky way, and could examine the relation between magnetic fields and the star-formation rate, down to very faint systems. A commensal HI survey is quite helpful, because the HI line profile informs redshift, gas state, dynamics of galaxies, and the mass ratio between HI and stars.

\begin{figure}[tbp]
\begin{center}
\FigureFile(80mm,80mm){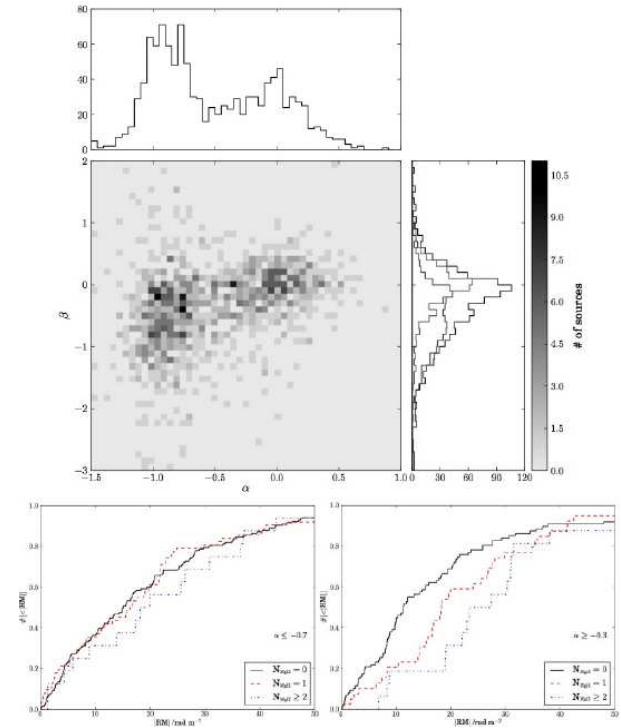}
\end{center}
\caption{
(Top panel) The relation between the total intensity spectrum index $\alpha$ and the polarization spectrum index $\beta$ for 951 sources \citep{2014ApJS..212...15F}, where the indices were derived from multiple observations between 400~MHz and 100~GHz. (Bottom two panels) The correlation between RM and the existence of foreground MgII absorption systems \citep{2014ApJ...795...63F}. The correlation depends on the type of sources ($\alpha$).
}
\label{f05}
\end{figure}

The third approach is to observe galaxies intervening the LOS toward polarized sources. This approach does not suffer from the bias that we look brighter galaxies. Combining with MgII optical line/imaging surveys, we could search the strength, growth time, and coherence length of galactic dynamo in disks and halos of normal galaxies, as well as beam covering fractions and spatial sizes of intervening galaxies, all as a function of redshift \citep{2008ApJ...676...70K, 2012ApJ...761..144B, 2013ApJ...772L..28B}. Figure~\ref{f05} shows the relation between the total intensity spectrum index $\alpha$ and the polarization spectrum index $\beta$ \citep{2014ApJS..212...15F}. There are two populations, an optically-thick AGN-core type ($\alpha\sim 0$) and an optically-thin radio-lobe type ($\alpha\sim -0.7$). The behavior of $\beta$ depends on $\alpha$, therefore, part of depolarization may take place at the source because $\alpha$ should depend only on the source nature. On the other hand, \citet{2014ApJ...795...63F} suggested that there is a clear correlation between RM and the existence of foreground MgII absorption systems for the AGN-core type (Fig. \ref{f05}). A large number of sources to be found with the SKA will allow us to test these features much more accurately.

{\it Related SKA Science Book 2015 Chapters; \citet{2015aska.confE.103G}, \citet{2015aska.confE.112S}, \citet{2015aska.confE.113T}.}

%%%%%%%%%%%%%%%%%%%%%%%%%%%%%%%%%%%%%%%%%%%%%%%
%%%%%%%%%%%%%%%%%%%%%%%%%%%%%%%%%%%%%%%%%%%%%%%
\subsection{AGN and Jets}
\label{mag.s2.ss5}

There are various unsolved problems on AGN jets, such as the acceleration mechanism, the acceleration spot, the composition of jets, and the influences from/to environment. A survey with the SKA1 will provide a catalog of nearby jets, and the catalog will allow us to tackle some key issues of AGN jets. 

For instance, several thousands of radio "quiet" AGNs will be discovered from the survey. We will investigate an essential difference in jet power between radio loud and quiet AGNs, using SKA-VLBI Band-3 and Band-5 observations. Such high resolution observations allow us to distinguish between a jet and star formation in radio quiet AGN systems. The star formation rate can be discussed, too. Various pc-scale jets will be observed and spatially resolved with SKA-VLBI Band-5. 

A survey with the sensitivity of 100~nJy in rms allows us to find the central massive black holes of normal galaxies located at up to $z\sim 8$, if the the AGN luminosity is about the Eddington luminosity. Such deep observations with the SKA1 could provide the information about co-evolution among the central black hole, galactic magnetic fields, and the host galaxy. Observing AGNs in wide ranges of fractional polarization and redshift enables us to make a catalog of magnetic-field structures as a function of time. 

\begin{figure}[tbp]
\begin{center}
\FigureFile(80mm,80mm){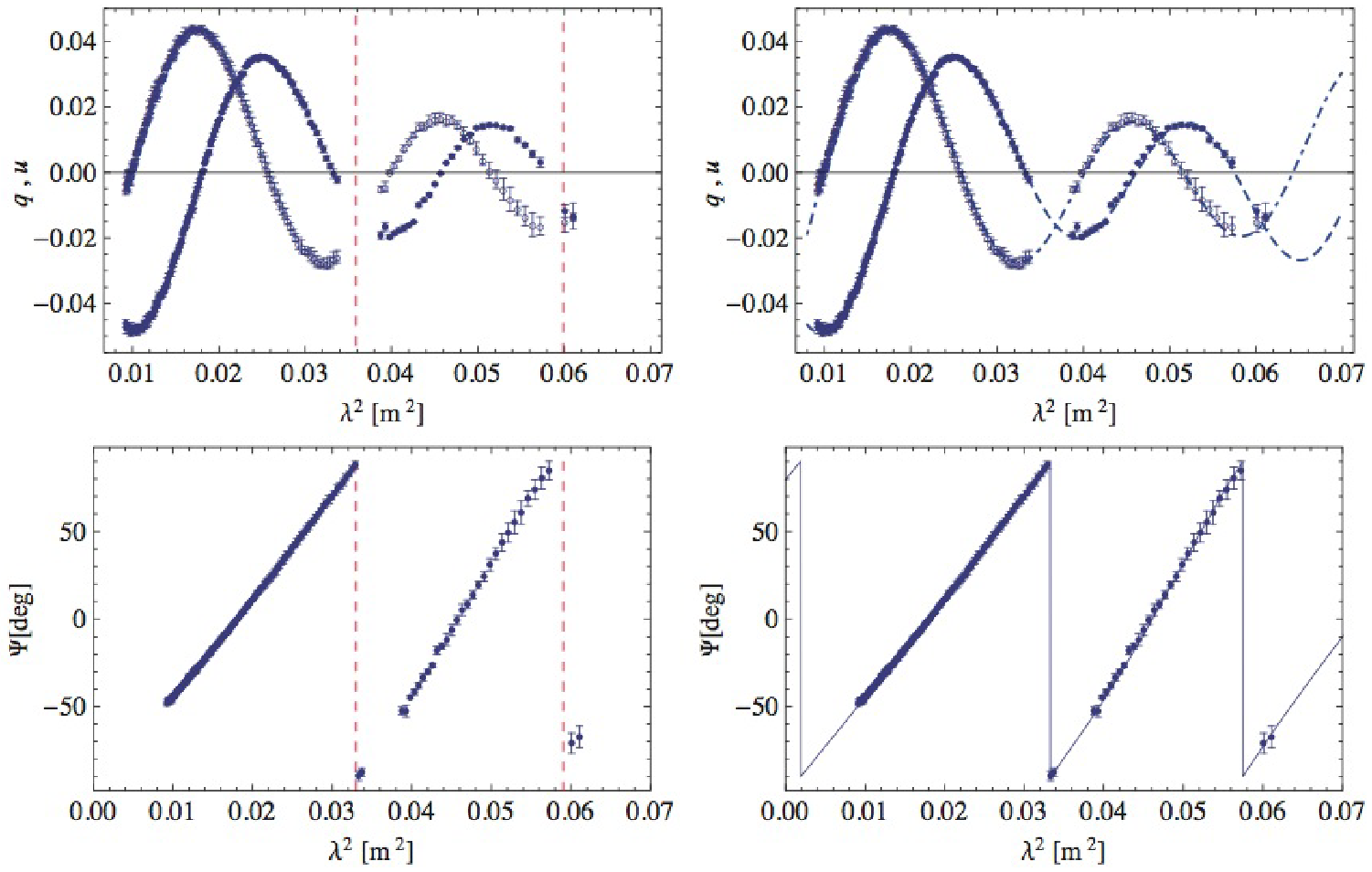}
\end{center}
\caption{
Polarization spectrum between 1.1~GHz  and 3.1~GHz of Quasar PKS B1610-771 \citep{2012MNRAS.421.3300O,2015aska.confE.103G}. Top panels show normalized Stokes q (=Q/I, open circles) and u (=U/I, filled circles) and the bottom panels show the polarization angle (filled circles) against the wavelength squared. Dashed lines in the left panels indicates the 350 MHz bandwidth centered at 1.4 GHz. Lines in the right panels show the best fit model with two RM components based on a full bandwidth data.
}
\label{f06}
\end{figure}

Wideband polarimetry is quite powerful for studying AGN jet structures. Figure~\ref{f06} introduces a result for distant quasar PKS B1610-771 \citep{2012MNRAS.421.3300O,2015aska.confE.103G}. If we employ narrow bandwidth data (350 MHz bandwidth centered at 1.4 GHz), we obtain a single RM component with +135 rad~m$^{-2}$. But with full bandwidth data (1.1 -- 3.1 GHz), they find that we obtain two polarized ``knots" with RMs of +107 rad~m$^{-2}$ and +79 rad~m$^{-2}$. This demonstrates that SKA's wideband studies could highlight structures of AGN, radio lobes, and surrounding gas even if they are spatially unresolved.

Observation of circular polarization will be able to give a direct measurement of magnetic field strength, a real magnetic flux propagating by a jet, and information about a composition of the jet (i.e. the positron--proton ratio). Thanks to high sensitivity and wide frequency coverage, SKA1 is an ideal device to observe the circular polarization, because Stokes V of AGN is typically weak ($\sim 0.1$ -- 1~\% of Stokes I) and it is hard to observe with the current facilities. If circular polarization is detected, SKA-VLBI should try to identify the emission region. We aim to determine physical parameters of the jet (strength, composition, magnetic flux) from the comparison with the observed data and theoretical models. 

{\it Related SKA Science Book 2015 Chapters; \citet{2015aska.confE..92J}, \citet{2015aska.confE..93A}, \citet{2015aska.confE.103G}, \citet{2015aska.confE.107L}, \citet{2015aska.confE.109P}, \citet{2015aska.confE.112S}, \citet{2015aska.confE.113T}.}

%%%%%%%%%%%%%%%%%%%%%%%%%%%%%%%%%%%%%%%%%%%%%%%
%%%%%%%%%%%%%%%%%%%%%%%%%%%%%%%%%%%%%%%%%%%%%%%
\subsection{Galaxy Clusters}
\label{mag.s2.ss6}

\begin{figure}[tbp]
\begin{center}
\FigureFile(75mm,75mm){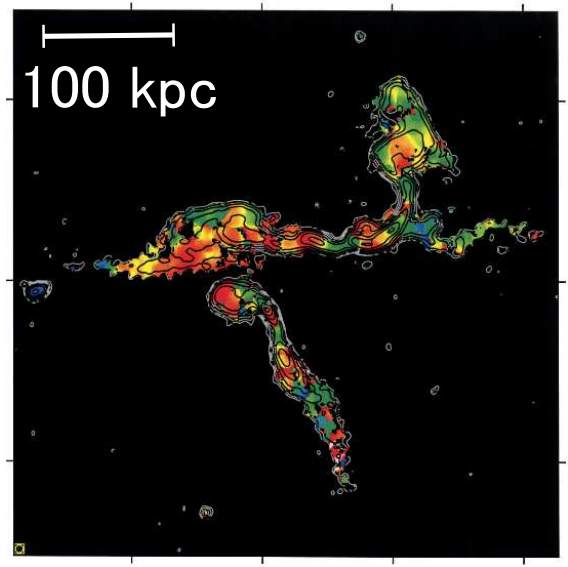}
\end{center}
\caption{
RM distribution of the radio galaxies in Abell 2382 \citep{2008A&A...483..699G}. Contours represent the radio emission and color represents the value of RM.
%%%(Top left) Radio halo in Abell 2219. (Top right) Radio relic in Abell 115. (Bottom left) Radio mini halo in Abell 2029. Contours represent the radio emissions and color represents the  X-ray emissions \citep{2012A&ARv..20...54F}. (Bottom right) 
}
\label{f07}
\end{figure}

It is expected that there are magnetic turbulence in the intracluster medium (ICM). At the same time, magnetic fields ordered along a specific direction may also exist due to compression of the gas by AGN outflows and cluster-merger shocks. Even Mpc-scale coherent magnetic fields might exist according to cosmological origins (\S \ref{mag.s3.ss11}). Clear evidence of magnetic fields is diffuse synchrotron emission called radio mini halo, radio halo, and radio relic, as well as Faraday rotation of background polarized sources; there are a lot of radio observations (e.g., \cite{2012A&ARv..20...54F}, Fig.~\ref{f07}). A brief summary of cluster magnetic fields is as follows. 

\begin{itemize}

\item{\bf Structure:} Micro-Gauss magnetic fields in the ICM have been reported. These spatial power spectra can be inferred from RM of back ground polarized sources, and so on. RM maps imply the Kolmogorov-like power spectrum \citep{2008A&A...483..699G}. A weak polarized intensity of radio halos implies significant power at small spatial scales.

\item{\bf Origin and evolution:} There are several hypotheses of the origin and amplification of cluster magnetic fields \citep{2000ApJ...541...88V,2008Sci...320..909R}, but those are not yet proven by any observation. There are only a few attempts of cosmological MHD simulations of cluster formation. 

\item{\bf Modelling:} 
Simulators of RM maps using (semi)-analytic models of the gas density and magnetic fields have been developed (e.g. FARADAY, \cite{2004A&A...424..429M}). Comparison between the simulated and observed data allows us to quantify cluster magnetic fields. A detailed radial profile of RM could relax the degeneracy of parameters in the magnetic-field models. Such semi-analytic simulations are also useful for planning SKA observations.

\end{itemize}

With SKA1's improved sensitivity, we will be able to obtain dense RM grids and study structures of cluster magnetic fields in detail. About three hundred polarized sources per square degree would be detected with the SKA1 at the sensitivity of 1~$\mu$Jy in rms at 1.4~GHz and the angular resolution of $1''.6$. If we observe a Coma-like cluster, $\sim 50$ polarized sources, which is 7 times larger than the VLA observations, would be detected. 

In the early operation ($\sim 50$~\% sensitivity) of the SKA1, the number of detected polarized sources are sufficient to study massive clusters above $10^{15}$ solar mass. They are, on the other hand, insufficient to study normal clusters and shock waves. Full SKA1 and SKA phase 2 (SKA2) are needed to address magnetic fields in normal clusters and shock waves. 

Information of the total and polarized intensity spectra would also be obtained from the improvement of the bandwidth. These information can be used to determine the energy spectrum of CRs. The SKA2 will get a 10 times better sensitivity and 20 times wider FoV. Improvement of the spatial resolution can reveal magnetic fields in smaller scales.

{\it Related SKA Science Book 2015 Chapters; \citet{2015aska.confE..95B}, \citet{2015aska.confE..100C}, \citet{2015aska.confE.101J}, \citet{2015aska.confE.104G}, \citet{2015aska.confE.105G}.}

%%%%%%%%%%%%%%%%%%%%%%%%%%%%%%%%%%%%%%%%%%%%%%%
%%%%%%%%%%%%%%%%%%%%%%%%%%%%%%%%%%%%%%%%%%%%%%%
\subsection{The Cosmic Web}
\label{mag.s2.ss7}

Galaxy filaments may be an unique site in which the information of primordial magnetic fields remains. If we find the intergalactic magnetic field (IGMF) in galaxy filaments, we may be able to test theoretical models of the generation and evolution of cosmological magnetic fields. The test would lead further discoveries related to the epoch of reionization, the first star formation, the cosmological recombination, even the physics of the early Universe. However, there have been few observations of the IGMF in galaxy filaments (e.g., \cite{2002NewA....7..249B}).

\begin{figure}[tbp]
\begin{center}
\FigureFile(80mm,80mm){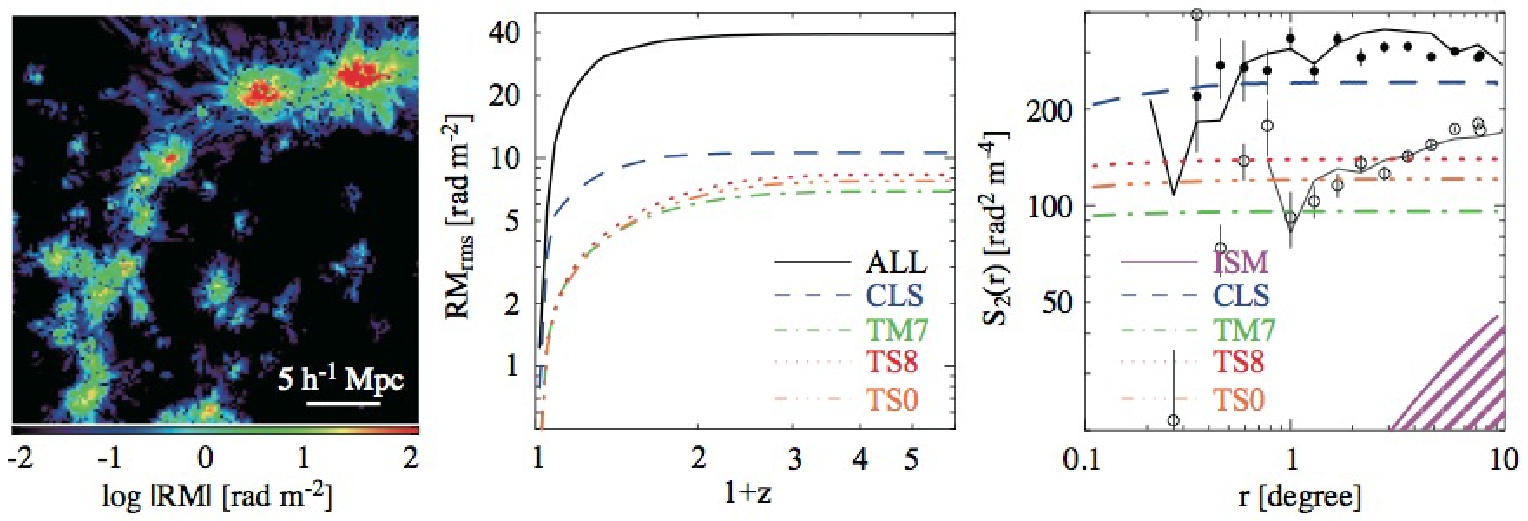}
\end{center}
\caption{
(Left) RM map of ($28~h^{-1}$ Mpc)$^2$ area in the local universe of depth of ($100~h^{-1}$ Mpc) \citep{2010ApJ...723..476A}. (Middle) The root-mean-square value of IGMF RMs integrated up to redshift of 5 \citep{2011ApJ...738..134A}. Colors indicate different methods of cluster subtraction. (Right) The second order structure function (SF) of RM. 
Black marks \citep{2010ApJ...714.1170M} and lines \citep{2011ApJ...726....4S} show the observed data for $\sim 900$ deg$^2$ toward the North (filled, thick) and South (open, thin) Galactic poles. The purple shaded region is a possible range of the Milky Way contribution based on the simulation \citep{2013ApJ...767..150A}.
}
\label{f08}
\end{figure}

The main reason of the lack of observations is that the IGMF is expected to be very weak. Cosmological simulations suggest that the IGMF in filaments may have $B\sim$1 -- 100 nG (e.g., \cite{2008Sci...320..909R}). Such a weak field gives a small RM. The rms value of RM through a single filament would be $\sim 1$ rad~m$^{-2}$ (Fig.~\ref{f08} left), while that through several filaments up to $z \sim {\rm a~few}$ would reach several rad~m$^{-2}$ (Fig.~\ref{f08} middle). The simulation predicts that the IGMF in filaments would induce RM with a flat second-order structure function (SF) with $\sim 100$ rad$^{2}$~m$^{-4}$ at angular scales of $r>0.1^\circ$ (Fig.~\ref{f08} right), while the Galactic magnetic fields should produce substantially smaller and steeper SF in small angular scales toward high galactic latitudes. A deep survey below the sensitivity 100~nJy in rms with SKA1-MID Band-2 will make it possible to figure out the SF at the scale $r>0.1^\circ$, and allow us to extract the RM of the IGMF in filaments. It is also expected very recently that fast radio burst (FRB) would be a new probe of the IGMF in filaments \citep{1602.03235}.

The information field theory would assist the detection of the IGMF in galaxy filaments  \citep{2009PhRvD..80j5005E}. The method statistically separates different extragalactic contributions, by combining RMs with the information about the large-scale structure of the cosmic web. Since the RM grid obtained with the SKA1 will be 300 -- 1000 times denser than the largest catalog available in the present, it will be possible to identify optical counterparts uniquely and hence the source redshift. An uncertainty in RM will be lower than $\sim 0.2$ rad~m$^{-2}$ by the observation of one hour and the resolution of $2''$ observation with SKA1-MID (350 -- 3050 MHz) which provides the polarization sensitivity below $1~\mu$Jy. These allow us to distinguish nearby components in Faraday spectra and improve statistics. Using the method, we could detect $\sim$~nG magnetic fields during the early operation ($\sim 50$~\% sensitivity) of the SKA1.

The predicted weak IGMF also make it difficult to observe synchrotron radiation from galaxy filaments. A sensitivity of $\sim 10~{\rm nJy/arcsec^2}$ at 1.4 GHz will be needed to detect the radiation, based on a few evidence of Mpc-scale diffuse emissions corresponding likely to galaxy filaments (e.g., \cite{2002NewA....7..249B}). This study is best achieved with SKA1-MID Band-2. A two hours observation with a 800 MHz bandwidth and a $3''$ resolution will provide the 0.08 $\mu$Jy/beam sensitivity, which allows us to detect synchrotron radiation from galaxy filaments at a 2.5 -- 6 $\sigma$ level.

{\it Related SKA Science Book 2015 Chapters; \citet{2015aska.confE..95B}, \citet{2015aska.confE..97V}, \citet{2015aska.confE.104G}, \citet{2015aska.confE.113T}, \citet{2015aska.confE.114V}.}

%%%%%%%%%%%%%%%%%%%%%%%%%%%%%%%%%%%%%%%%%%%%%%%
%%%%%%%%%%%%%%%%%%%%%%%%%%%%%%%%%%%%%%%%%%%%%%%
%%%%%%%%%%%%%%%%%%%%%%%%%%%%%%%%%%%%%%%%%%%%%%%
\section{Japanese SKA Magnetism Science}
\label{mag.s3}

For our feasibility verification below, we have referred the SKA1 Level 0 Science Requirements document (Date 2015-10-28). This has incorporated the re-baseline of the SKA1. Although the re-baseline decision does not include MID Bands 3, 4, and 5+, we have considered them as possible extensions of the SKA1. We have employed the 100 kHz point-source sensitivities of 184, 132, 66, 83, 100, and 132 $\mu$Jy~hr$^{-1/2}$ for SKA1 LOW, MID Bands 1, 2, 3, 4, and 5, respectively. We have ignored penalties of radio frequency interferences (RFIs).

Our scientific objectives and use cases introduced below are not official ones organized by the SKAO. We acknowledge SKA's policy on open co-development of sciences and notional key science projects, and hence our proposals do not insist on any Japan's priority. We also acknowledge international SKA SWGs, since some of our proposals are inspired gratefully by the ones developed by them.

%%%%%%%%%%%%%%%%%%%%%%%%%%%%%%%%%%%%%%%%%%%%%%%
%%%%%%%%%%%%%%%%%%%%%%%%%%%%%%%%%%%%%%%%%%%%%%%
\subsection{Pulsar RMs and the Galactic Magnetic Field Structure}
\label{mag.s3.ss1}

{\it This science objective and its use case, SKAJP-MAG-1 (ISM), are described in the SKA-JP Pulsar Science Chapter.}

%%%%%%%%%%%%%%%%%%%%%%%%%%%%%%%%%%%%%%%%%%%%%%%
%%%%%%%%%%%%%%%%%%%%%%%%%%%%%%%%%%%%%%%%%%%%%%%
\subsection{Unveiling Zebra Pattern in High Galactic RM Sky}
\label{mag.s3.ss2}

%%%%%%%%%%%%%%%%%%%%%%%%%%%%%%%%%%%%%%%%%%%%%%%
\subsubsection{Objective}
\label{mag.s3.ss2.sss1}

The origin and evolution of the large-scale magnetic field in the Milky Way is one of the most important subjects in SKA's cosmic magnetism. Using numerical simulation, we will reveal the importance of CRs that exist in the interstellar medium. The CRs strongly affect the magnetic buoyant instability, such as Parker instability, which is believed to be coupled with the galactic dynamo.

The numerical simulation of the Parker instability with CRs is first studied by \citet{2004ApJ...607..828K}, and followed by \citet{2014JPSCP.1.015105} under the situation of local galactic disks. We will study the galactic dynamo using numerical simulations of global galactic disks like \citet{2006ApJ...641..862N} and include CRs in the simulations. The numerical simulations of galactic dynamo with CRs was studied by \citet{2009ApJ...706L.155H}. However, the physical process of the galactic dynamo and the importance of CRs have not been clear yet. Our numerical simulations should be cooperated with observations as well as analytical theories.

\begin{figure}[tbp]
\begin{center}
\FigureFile(80mm,80mm){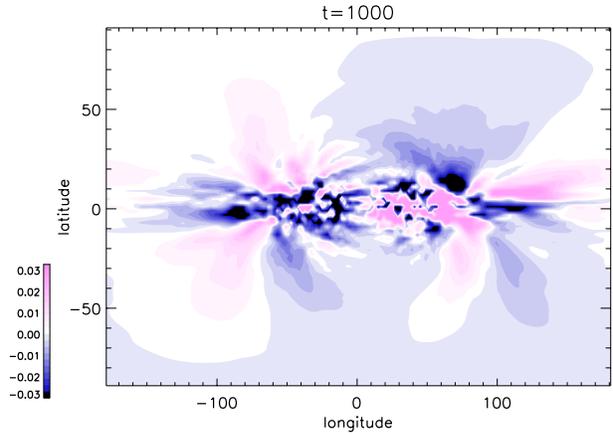}
\end{center}
\caption{
The zebra pattern in a high galactic RM sky predicted by the MHD simulation of galactic gaseous disks \citep{2013ApJ...764...81M}.
}
\label{f09}
\end{figure}

We propose the observations of RMs at high galactic latitudes. \citet{2013ApJ...767..150A} showed that the observed north and south anti-symmetric structure of magnetic fields cannot be explained by randomness of turbulent magnetic fields. They claimed that the origin of the anti-symmetric structure would be caused by the Parker instability. \citet{2013ApJ...764...81M} performed the numerical simulations of galactic dynamo in the global disk, and showed that the north and south magnetic structures are periodically exchanged because of the magnetic buoyancy. The model predicts a zebra pattern of RM at high galactic latitudes (Fig. \ref{f09}). The observational results should be compared with that of the numerical simulation. We also study the effect of CRs on the anti-symmetric structure using the similar numerical simulations of \citet{2013ApJ...764...81M} by including CRs into the code.

%%%%%%%%%%%%%%%%%%%%%%%%%%%%%%%%%%%%%%%%%%%%%%%
\subsubsection{Use Case: SKAJP-MAG-2 (MWG)}
\label{mag.s3.ss2.sss2}

\begin{table}
\tbl{Key Requirements of SKAJP-MAG-2 (MWG)}{%
\begin{tabular}{llll}
\hline
Central frequency (Band) & 200 MHz (LOW) \\
Spectral coverage and resolution & 300 MHz, 1 MHz \\
Angular resolution (baseline) & $50''$ (7 km)\\
Image rms (I, P) and bandwidth & (1440, 1440) $\mu$Jy/beam, 1 MHz\\
FoV and pointings & 6.3 deg$^2$, 548$^\dagger$ (Mosaicking) \\
Total on-source time & 67 hours\\
All-sky Band-2 & Desirable$^\ddagger$ \\
\hline
\end{tabular}}\label{tab:SKAJP-MAG-2}
\begin{tabnote}
$^\dagger$ $|b|>30^\circ$ and $|l| < 30^\circ$ (total $\pi/3$ sr = 3438 deg$^2$).\\
$^\ddagger$ Use to subtract point sources and perform Faraday Tomography.
\end{tabnote}
\end{table}

As described above, there can be halo toroidal fields induced by the magneto-rotational instability and the Parker instability. Such toroidal fields are expected to remain, showing a zebra pattern (with a scale of $\sim 5^\circ$) of Stokes I and RM at high galactic latitudes \citep{2013ApJ...764...81M,morita16}. We try to discover the pattern in a Stokes I map for low frequencies. Furthermore, with a full band polarimetric data, we attempt to study sub-patterns ($\sim$ degree scales) in the RM map by means of Faraday synthesis, resolving LOS structures on some levels. We finally perform Fourier analyses of Stokes I and RM maps and compare the power spectra with those of our theoretical prediction. This study will constrain real mechanisms of the origin and evolution of Milky Way magnetic fields.

The standard observing procedure can be adopted. More specifically, the project can be divided into many scheduling blocks with an appropriate mosaicking strategy. Each block may start/end observations of band-pass calibrators and consists of observations of targets and gain/phase calibrators. Nighttime observation is preferable to avoid possible RFIs seen in the daytime. If daytime, we will avoid observations within 10 degree from the Sun. We should note that the calibration stability in time should be taken into consideration very much in low frequency observations. 

A minimum antenna baseline of $\sim 40$~m at zenith will provide the maximum angular scale of $\sim 5^\circ$ at 150~MHz, and this specification could satisfactorily capture the zebra pattern we are looking for. A planning concept of sub-station operation is quite promising for this use case and will allow us to clarify the error of the missing flux. 

We will utilize data of the all-sky Band-2 survey. The centimeter data would be useful to subtract point sources. We also attempt to extract the Milky Way foreground component, by incorporating the centimeter data into Faraday tomography. For LOW data only, specifications of Faraday Synthesis are as follows: the Faraday depth resolution is 0.098 rad~m$^{-2}$ (50 -- 350~MHz), the maximum observable Faraday depth is 833.7 rad~m$^{-2}$ (349 -- 350~MHz), and the maximum observable Faraday thickness is 4.28 rad~m$^{-2}$ (350~MHz).

%%%%%%%%%%%%%%%%%%%%%%%%%%%%%%%%%%%%%%%%%%%%%%%
%%%%%%%%%%%%%%%%%%%%%%%%%%%%%%%%%%%%%%%%%%%%%%%
\subsection{Classification of Polarized Pattern in Nearby Edge-on Galaxies}
\label{mag.s3.ss3}

%%%%%%%%%%%%%%%%%%%%%%%%%%%%%%%%%%%%%%%%%%%%%%%
\subsubsection{Objective}
\label{mag.s3.ss3.sss1}

Galactic magnetic fields are thought to play important roles in star formation, interstellar turbulence, activity of galactic center, and galactic wind which transports magnetized plasma from disk to halo. Some scenarios of the origin of ordered magnetic fields have been proposed; for example, a primordial-origin model for composite configurations of global magnetic fields \citep{2010PASJ...62.1191S} and a galactic dynamo model \citep{1986ARA&A..24..459S}. Disk ordered magnetic fields are mainly classified into Axis Symmetric Spiral (ASS) and Bi Symmetric Spiral (BSS) (see .e.g., \cite{1986ARA&A..24..459S}). In addition, \citet{2011MNRAS.412.2396F} reported composite structures: ASS in disk and BSS in halo for M51. Otherwise, edge-on view of ordered fields show X-shape structure (e.g. \cite{2013A&A...560A..42M}).

Although there are many attempts of morphological classification of galactic magnetic fields, it is still difficult to compare theoretical models with observations due to a lack of galaxy samples. For example, one of the current largest galaxy surveys WSRT-SINGS \citep{2007A&A...461..455B,2009A&A...503..409H,2010A&A...514A..42B} presented galaxies including only 35 edge-on nearby spiral galaxies, which is not enough to perform statistical analysis by classifying samples into some categories based on the morphology. We also need to improve the method of morphological classification. Up to now, several tens of face-on galaxies have ever been classified into ASS or BSS, though the classification is still under discussion \citep{anraku16}. 

Given this situation, we conclude that a systematic and extensive survey of nearby galaxies is of crucially importance. Therefore, we propose a multi frequency polarimetric survey for polarized synchrotron radiation and RM of nearby galaxies. We will carry out morphological classification of ordered magnetic fields parallel to and perpendicular to the disk. To search for diversity and universality of magnetic-field morphology, we need a few hundred samples of external galaxies. We must accomplish this goal by SKA's high-sensitivity mission.

In advance of our survey, we have already started to analyze archival data of polarimetric observations and are developing a new method of morphological classification of ordered magnetic fields in external galaxies \citep{anraku16}. Moreover, we are examining the best observing strategy based on Faraday depolarization and Faraday tomography. For instance, using data of MHD simulations, \citet{2014ApJ...792...51I} showed that we have to carefully consider the beam size in order to avoid complexities in features of depolarization and Faraday tomography. Although depolarization and Faraday tomography have the potential to deproject the LOS toroidal and poloidal components in the disk and halo, they do not work effectively without the consideration. 

Finally, our morphological classification can naturally extend to distant galaxies, once the supreme sensitivity and ultra-high resolution of the SKA2 are given. Such extension may allow us to discover the cosmological co-evolution between galactic magnetic fields and galaxies. We can collaborate with Japanese "galactic evolution" and "interstellar medium" SWGs in this project.

%%%%%%%%%%%%%%%%%%%%%%%%%%%%%%%%%%%%%%%%%%%%%%%
\subsubsection{Use Case: SKAJP-MAG-3 (NGL)}
\label{mag.s3.ss3.sss2}

\begin{table}
\tbl{Key Requirements of SKAJP-MAG-3 (NGL)}{%
\begin{tabular}{llll}
\hline
Central frequency (Band) & 3990 MHz (Band 4) \\
Spectral coverage and resolution & 2380 MHz, 100 MHz \\
Angular resolution (baseline) & $2.''5$ (6 km)\\
Image rms (I, P) and bandwidth & (2.0, 2.0) $\mu$Jy/beam, 100 MHz\\
FoV and pointings  & 0.086 deg$^2$, 200 (nearby galaxies) \\
Total on-source time & 500 hours\\
All-sky Band-2 & Essential$^\dagger$ \\
\hline
\end{tabular}}\label{tab:SKAJP-MAG-3}
\begin{tabnote}
$^\dagger$ Use to select sources, check depolarization, and perform Faraday tomography.
\end{tabnote}
\end{table}

We propose observations of the nearby ``edge-on" galaxies with a high sensitivity and a high spatial resolution. In order to classify polarized pattern into some categories, we will need 200 targets. We will select targets from the Tully's bright galaxy catalog (which includes more than 200 galaxies and 12 radio counterparts have been detected), where we will discard the galaxies whose inclination angle $<60^\circ$ or declination $>0^\circ$. SKA1-MID Band-4 is the best choice because (i) we can gain good spatial resolution (iii) we can avoid too-strong depolarization and  (ii) synchrotron emission can be not too weak. Using NGC253 as a typical flux density of targets for our sensitivity calculation, we request 500 hours observations on source, where we require the $2.''5$ angular resolution corresponding to a spatial resolution of 100~pc at $z\sim 0.001$.

The standard observing procedure (scheduling of calibrator and target observations) and standard data reduction (calibration, RFI flagging, and imaging) can be adopted. Each target can be observed with a single pointing, since a typical size of nearby galaxy, $1'$ -- $10'$, can be covered with a FoV of $16'.5$ (SKA1-MID Band-4). This is follow-up observations of the all-sky Band-2 survey. We suppose that we have nearby edge-on galaxy candidates in the survey catalog. A table of band-pass and gain/phase calibrators must be available for Band-4.

%%%%%%%%%%%%%%%%%%%%%%%%%%%%%%%%%%%%%%%%%%%%%%%
%%%%%%%%%%%%%%%%%%%%%%%%%%%%%%%%%%%%%%%%%%%%%%%
\subsection{Cosmic Evolution of DIG's Magnetic Fields}
\label{mag.s3.ss4}

%%%%%%%%%%%%%%%%%%%%%%%%%%%%%%%%%%%%%%%%%%%%%%%
\subsubsection{Objective}
\label{mag.s3.ss4.sss1}

Observations of galaxies over redshifts are a straightforward way to elucidate the real evolution history of galactic magnetic fields. As introduced in \S\ref{mag.s2.ss4}, depolarization properties of background emission allow us to study depolarizing intervening galaxies (DIGs). This approach fits well with the Japanese focus on depolarization and Faraday tomography.

Optical (MgII absorption) observations indicate the existence of absorbers in front of about a half of SDSS galaxies including polarized sources such as radio galaxies and quasars \citep{2013ApJ...770..130Z}. The absorbers are likely DIGs, which can alter appearance of background polarization \citep{2012ApJ...761..144B,2014ApJ...795...63F}. With SKA's wideband capability, we could extend the previous works \citep{2014ApJS..212...15F,2014ApJ...795...63F} dramatically. Increase of the number of samples allows us to study universality and diversity of DIG's properties, and allow us to figure out the cosmological evolution of galactic magnetic fields. The SKA is an unique facility which makes these observations possible.

\citet{Akahori2016} tested depolarization caused by DIGs, using a model of the Milky Way. Their preliminary results are as follows. Both the global and local magnetic fields can contribute to depolarization, and its significance depends highly on observational specifications such as the observing frequency, the observing beam size, and the pointing center. The Burn law is not satisfied, because RM distribution within the beam does not follow the Gaussian. DIG's contribution to the observed RM decreases significantly as the DIG's redshift increases. Combining SKA observations with these theoretical works are quite powerful to extract more information about magnetic fields and turbulence.

As future works for the DIG simulations, it would be important to consider the population of galaxies such as the mass function and morphology. Also, application of Faraday tomography and to learn about Faraday spectrum would be important. Contribution of DIGs to observables becomes an uncertainty on the studies of other ``bright" components such as the source itself and the Milky Way. However, once we understand the properties caused by the DIG, such as depolarization features and Faraday spectra, we could judge the existence of DIGs and could extract polarized sources toward which there is no DIGs. Such a source selection is vital for the search of the IGMF. Therefore, it is fruitful to investigate the properties of DIGs.

%%%%%%%%%%%%%%%%%%%%%%%%%%%%%%%%%%%%%%%%%%%%%%%
\subsubsection{Use Case: SKAJP-MAG-4 (DGL)}
\label{mag.s3.ss4.sss2}

\begin{table}
\tbl{Key Requirements of SKAJP-MAG-4 (DGL)}{%
\begin{tabular}{llll}
\hline
Central frequency (Band) & 700 MHz (Band-1) \\
Spectral coverage and resolution & 770 MHz, 1 MHz \\
Angular resolution (baseline) & $2''$ (45 km)\\
Image rms (I, P) and bandwidth & (4, 4) $\mu$Jy/beam, 300 MHz\\
FoV and pointings & 2.8 deg$^2$, 500 (polarized sources) \\
Total on-source time & 182 hours\\
All-sky Band-2 & Essential$^\dagger$ \\
\hline
\end{tabular}}\label{tab:SKAJP-MAG-4}
\begin{tabnote}
$^\dagger$ Use to select sources, check depolarization, and perform Faraday Tomography.
\end{tabnote}
\end{table}

Since depolarization depends significantly on both regular and turbulent magnetic fields as well as observational specifications (as described, \cite{Akahori2016}), a detailed analysis with large samples is necessary to extract the cosmic evolution of magnetic fields in DIGs from observations of background sources. Wideband polarimetry is essential for this science, and the all-sky Band-2 survey is insufficient in terms of its frequency coverage. The Band-1 follow-up is necessary to capture rich depolarization properties caused by DIGs and to make Faraday tomography much promising. Therefore, we propose a follow-up project of the all-sky Band-2 survey. We suppose that we have a polarization catalog of the Band-2 survey and an all-sky catalog of MgII absorber systems. Then, from these catalogs, we will select 500 DIGs distributing evenly in redshift $z\sim 0$ -- 1 with a bin of $\delta z=0.1$ (50 DIGs in each redshift bin). 

Table~\ref{tab:SKAJP-MAG-4} summarizes key requirements of this use case. In order to perform systematic analysis combining with the Band-2 data, the spatial resolution (the imaged pixel size) should be the same as that of Band-2, i.e. $2''$. We require the image rms level of 4 $\mu$Jy/beam with 300 MHz bandwidth. This sensitivity would provide a comparable detection limit to the Band-2 survey, for AGN-core type sources ($\alpha \sim 0$), and a better limit for radio-lobe type sources ($\alpha \sim -0.7$) (see \S\ref{mag.s2.ss4}). We finally request 182 hours on-source observing time. 

The standard observing procedure can be adopted. We may need a careful scheduling to avoid penalties of Band-1 RFIs which may depend on time (day/night) and season (summer/winter). The project could be commensal with HI, cosmology, and continuum survey. In this case, we need to consult with these survey teams about observing strategy, too. The standard calibration, optimized flagging, and imaging are performed. The calibrated visibilities are imaged using a 300 MHz simultaneous bandwidth with robust weighting. Source finding is performed on Stokes-I basis with the images. Once detected, the bandwidth is recursively divided by half down to the 1 MHz channel or down to the detection limit so as to gain spectroscopic data. Here, a 1 MHz channel provides a maximum Faraday depth of about 5000 rad~m$^{-2}$ at 700 MHz. For I, Q, and U of each source, Faraday synthesis is carried out followed by Faraday RM clean with standard parameters. In addition, for each source, Q, and U are fitted with several FDF models by means of an MCMC method, where the result of RM clean is used as an initial guess.

%%%%%%%%%%%%%%%%%%%%%%%%%%%%%%%%%%%%%%%%%%%%%%%
%%%%%%%%%%%%%%%%%%%%%%%%%%%%%%%%%%%%%%%%%%%%%%%
\subsection{Typical magnetic-field structure of FR galaxies}
\label{mag.s3.ss5}

%%%%%%%%%%%%%%%%%%%%%%%%%%%%%%%%%%%%%%%%%%%%%%%
\subsubsection{Objective}
\label{mag.s3.ss5.sss1}

We, Japanese researchers, have performed various numerical simulations related to jets, such as jet formation from an accretion disk, propagation of jets, and so on, using supercomputers (e.g., \cite{1985PASJ...37...31S,1996ApJ...461..115M,2006ApJ...652.1059N}). We learned jet mechanisms closely from the simulations, and can perform new jet simulations. This is Japan's advantage for AGN jet science. Specifically, we already have (and can newly generate) numerical data which could compare to the data of future SKA observations. From the comparison, we will address various unsolved problems on AGN jets such as the acceleration mechanism, the acceleration spot, the composition, and the correlation with environment. 

For example, by numerical simulation, we have demonstrated the basic jet structure: vertical magnetic fields dominate around the jet axis while toroidal magnetic fields surround the jet. In other word, jets consist of two layers -- an inner high speed region and an outer slow outflow, called ``spine" and ``sheath", respectively. Such numerical simulations were calculated under various initial conditions, such as initial configuration of magnetic fields or distribution of ambient matter. From the simulations, we learned that the structure of surrounding magnetic fields, particularly near a front edge of the jet, depends on the jet launch mechanism. The primary jet launch mechanism is then related to the initial structure (e.g., toroidal, poloidal, and vertical) of magnetic fields in an accretion disk. These mean that we can get information about the initial condition of jet launch, from the observed, evolved jet structure. 

We, therefore, examine the structure of magnetic fields around the jet with the SKA. From comparison between observational data and numerical results, we could get information about the initial condition of jet launch, e.g. initial magnetic fields in an accretion disk. Such information is essential for discussing several unknown processes such as how to launch jets, how to collimate jets, where is the accelerating point, and so on. In order to achieve this work, we need to observe jet structure, and hence we choose FR-type galaxies for our targets. 

We have already started observational visualization (Stokes parameters and RM) of numerical data \citep{2004ApJ...600...88U,2004ApJ...608..119K,morita16}. We will obtain the best-fit model of initial magnetic-field structure and be able to pull up information of magnetic fields inside an accretion disk. We will also prepare templates of Faraday spectra obtained from numerical results. From comparision between the simulated Faraday spectra and observed spectra, we will try to understand typical Faraday spectrum of FR galaxies, especially what is the difference of Faraday spectra between FRI and FRII.

%%%%%%%%%%%%%%%%%%%%%%%%%%%%%%%%%%%%%%%%%%%%%%%
\subsubsection{Use Case: SKAJP-MAG-5 (JET)}
\label{mag.s3.ss5.sss2}

\begin{table}
\tbl{Key Requirements of SKAJP-MAG-5 (JET)}{%
\begin{tabular}{llll}
\hline
Central frequency (Band) & 1355 MHz (Band-2) \\
Spectral coverage and resolution & 810 MHz, 10 MHz\\
Angular resolution (baseline) & $0''.5$ (92 km)\\
Image rms (I, P) and bandwidth & (0.43, 0.43) $\mu$Jy/beam, 10 MHz\\
FoV and pointings & 0.75 deg$^2$, 18 (FRI galaxies)\\
Total on-source time & 434 hours\\
All-sky Band-2 & --  \\
\hline
Central frequency (Band) & 1355 MHz (Band-2) \\
Spectral coverage and resolution & 810 MHz, 10 MHz\\
Angular resolution (baseline) & $2''$ (23 km)\\
Image rms (I, P) & (0.27, 0.27) $\mu$Jy/beam\\
FoV and pointings & 0.75 deg$^2$, 12 (FRII galaxies)\\
Total on-source time & 734 hours\\
All-sky Band-2 & --  \\
\hline
\end{tabular}}\label{tab:SKAJP-MAG-5}
\end{table}

We propose observations of nearby FRI and FRII galaxies. We try to resolve sub arcsec- to arcsec-scale jet structures at $\sim$ GHz with the SKA1. Such an angular resolution is about 10 times better than those of SKA precursors/pathfinders. We emphasize that high-resolution observations can advance not only study of spatial structures but also study of depolarization and Faraday tomography. This is because relatively young jets and robes can possess very large RMs, which lead serious beam depolarization and make it difficult to obtain meaningful results if the beam size is large. Evaluation of the influence of beam depolarization will be a key to correctly extract physical quantities from comparison between observational data and numerical ones. Fortunately, we expect that the number of observed FR galaxies will increase dramatically in the SKA era, we will be able to examine the influence with many samples statistically.

From 71 FRI and 466 FRII galaxies in the combined NVSS-FIRST sample (CoNFIG), we choose 18 FRI and 12 FRII galaxies. Our selection criteria are (i) the spectral index is known, (ii) the Galactic longitude less that $+44^\circ$, (iii) the 1.4 GHz flux larger than $\sim 1$~Jy, and (iv) the redshift less than 0.1.

The standard observing procedure (scheduling of calibrator and target observations) and standard data reduction (calibration, RFI flagging, and imaging) can be adopted. Each of our 30 targets is observed with a single pointing, since the nearby FR galaxies can be covered with a FOV of $48'.5$ (SKA1-MID Band-2). We are proposing deeper observations than the all-sky Band-2 survey. Data of the all-sky Band-2 survey will be helpful for us to tune up our observing strategy, e.g. we can adjust an observing time for each target.

We note that it is difficult to spatially resolve magnetic-field structures in an accretion disk. Therefore, we will refer X-ray observations and/or numerical simulations, when we will need to specify properties of the accretion disk.

%%%%%%%%%%%%%%%%%%%%%%%%%%%%%%%%%%%%%%%%%%%%%%%
%%%%%%%%%%%%%%%%%%%%%%%%%%%%%%%%%%%%%%%%%%%%%%%
\subsection{Observations of 20 Radio Mini-Halos}
\label{mag.s3.ss6}

%%%%%%%%%%%%%%%%%%%%%%%%%%%%%%%%%%%%%%%%%%%%%%%
\subsubsection{Objective}
\label{mag.s3.ss6.sss1}

Radio mini-halos are diffuse radio synchrotron emissions observed around the cores of galaxy clusters. Contrary to cluster-scale radio halos, the mini-halos are often observed in non-merging clusters \citep{2012A&ARv..20...54F,2014ApJ...781....9G}. This suggests that the origin of the mini-halos is different from that of cluster-scale radio halos. Since mini-halos are generally dim, the unprecedented sensitivity of the SKA is expected to play an important role to reveal the origin of the mini-halos. For example, the mini-halos have been found in only a fraction of clusters. We do not know whether mini-halos are too dim to detect in most clusters or they are totally absent in them. SKA observations will be able to distinguish the two possibilities.

Japan has an advantage in the research of magnetic fields and CRs in clusters, because intensive multi-wavelength observations have been planned. Thus, we propose multi-wavelength observations of the mini-halos using the SKA and other telescopes. First, the Japanese X-ray satellite with international collaboration, Hitomi (Astro-H), has a superb spectral resolution, and it can detect the Doppler-shift of hot ICM with the resolution of $<100\rm km\: s^{-1}$ \citep{2014arXiv1412.1176K}. This will enable us to measure turbulence in the ICM \citep{2015arXiv150702730O,2015arXiv150804426Z}.

Although the sensitivity of Hitomi is not too high, turbulence in cluster cores should be detected because the cores are bright enough. The information of the turbulence can be compared with detailed radio observations of mini-halos obtained with the SKA. For example, if a synchrotron-emitting region is found to be turbulent, primary CR electrons accelerated in the turbulence are likely to be responsible for the synchrotron emission \citep{2013ApJ...762...78Z}. Moreover, by comparing the distribution of turbulence, synchrotron emissions, and Faraday rotation, we may be able to discuss the amplification of magnetic fields in the turbulence. On the other hand, if there is no correlation between turbulence and synchrotron emissions, it may mean that the emissions come from secondary CR electrons created via interaction between CR protons and ICM protons (e.g., \cite{2012ApJ...746...53F}). Combined analysis of radio spectra obtained with the SKA and hard X-ray spectra obtained with Hitomi will reveal the energy spectra of CRs and magnetic field strength in the mini-halos.

Besides, optical weak-lensing observations with Hyper Suprime-Cam (HSC) on the Subaru telescope will tell us the distribution of matter in clusters \citep{2015ApJ...807...22M}. If a cluster has an irregular matter distribution, it means that the cluster is in the merging process and turbulence is likely to be excited in the ICM. If mini-halos are produced by secondary electrons, gamma-rays associated with the proton-proton interaction may be observed with Cherenkov Telescope Array (CTA) \citep{2011ExA....32..193A}. By comparing the above observations with SKA observations, our knowledge about mini-halos will be improved dramatically.

%%%%%%%%%%%%%%%%%%%%%%%%%%%%%%%%%%%%%%%%%%%%%%%
\subsubsection{Use Case: SKAJP-MAG-6 (CMH)}
\label{mag.s3.ss6.sss2}

\begin{figure}[tbp]
\begin{center}
\FigureFile(80mm,80mm){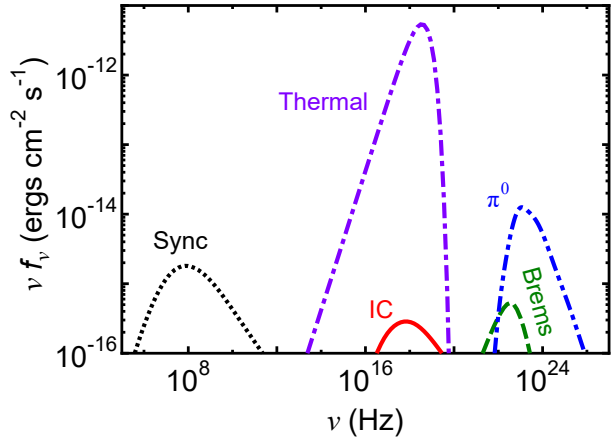}
\end{center}
\caption{
Predicted Broad-band spectra of RX~J1347--1145 based on the secondary cosmic-ray scenario by \citet{2013MNRAS.428..599F}. Synchrotron radiation (dotted line), inverse Compton scattering off cosmic microwave background (solid line), and non-thermal bremsstrahlung (dashed line) are of the secondary electrons. The $\pi^0$-decay gamma-rays are shown by the two-dot-dashed line. For comparison, the thermal bremsstrahlung from the ICM is shown by the dot-dashed line. Redshift has been corrected.
}\label{f10}
\end{figure}

\begin{table}
\tbl{Key Requirements of SKAJP-MAG-6 (CMH)}{%
\begin{tabular}{llll}
\hline
Central frequency (Band) & 200 MHz (LOW) \\
Spectral coverage and resolution & 300 MHz, 10 MHz\\
Angular resolution (baseline) & $12''$ (25 km)\\
Image rms (I, P) and bandwidth & (51, 51) $\mu$Jy/beam, 10 MHz\\
FoV and pointings & 6.3 deg$^2$, 20 (mini halos)\\
Total on-source time & 332 hours\\
All-sky Band-2 & Essential$^\dagger$ \\
\hline
Central frequency (Band) & 1355 MHz (Band-2) \\
Spectral coverage and resolution & 810 MHz, 10 MHz\\
Angular resolution (baseline) & $25''$ (2.1 km)\\
Image rms (I, P) and bandwidth & (22, 22) $\mu$Jy/beam, 10 MHz\\
FoV and pointings & 0.75 deg$^2$, 20 (mini halos)\\
Total on-source time & 164 hours\\
All-sky Band-2 & Essential$^\dagger$ \\
\hline
\end{tabular}}\label{tab:SKAJP-MAG-6}
\begin{tabnote}
$^\dagger$ Use to select sources.
\end{tabnote}
\end{table}

In order to identify the origin, we observe radio mini-halos in galaxy clusters. We study spatial distributions, spectra, and polarization properties of the mini-halos, and compare them with theoretical predictions (e.g., \cite{2013MNRAS.428..599F}, Fig.~\ref{f10}). If mini-halos are generated by secondary electrons, the radio profiles should be well correlated with X-ray profiles of the ICM \citep{2013MNRAS.428..599F}. Otherwise, if mini-halos are generated by primary electrons, the correlation between the radio profiles and the X-ray profiles should be weak, while the radio emission should be correlated with turbulence \citep{2013ApJ...762...78Z}. Spectra can be used to estimate the age of CR electrons. Polarization properties such as the polarization angle and polarization degree can be used to find the configuration of magnetic fields.

We suppose that we have mini-halo candidates in the all-sky Band-2 survey catalog and we select 20 best targets from the candidates. The project (each LOW and MID2) can be divided into 20 scheduling blocks for the 20 targets \citep{2014ApJ...781....9G} and the standard observing procedure can be adopted. A table of band-pass and gain/phase calibrators must be available. Each target is observed with a single pointing, since a typical size of mini-halos, 1 arc minute, can be covered with a FoV of 6.28 (LOW) and 0.75 (MID2) square degrees. We have adopted a radio mini-halo in RX J1347.5-1145 as a representative target, and have estimated a total observing time using the flux density of the mini-halo. The total observing time hence varies for actual targets selected.

%%%%%%%%%%%%%%%%%%%%%%%%%%%%%%%%%%%%%%%%%%%%%%%
%%%%%%%%%%%%%%%%%%%%%%%%%%%%%%%%%%%%%%%%%%%%%%%
\subsection{Deep Spectral Index Mapping of 6 Radio Halos} \label{mag.s3.ss7}

%%%%%%%%%%%%%%%%%%%%%%%%%%%%%%%%%%%%%%%%%%%%%%%
\subsubsection{Objective}
\label{mag.s3.ss7.sss1}

Radio halos are known as diffuse synchrotron emission of Mpc extent. The surface brightness is low ($\sim 0.1$ -- 1~$\mu$Jy/arcsec$^2$ at $\nu = 1.4$ GHz), and the spectral index is steep ($\alpha \sim 1$ -- 2) \citep{2012A&ARv..20...54F}. The magnetic field strength is estimated to be $\sim 0.1$ -- 1~$\mu$G. It is considered that fluctuations seen in the surface brightness reflect the structure of magnetic fields. While spatial distributions of radio and X-ray emission tend to be similar each other in giant, regular radio halos, such association is not often seen in small, irregular radio halos. The difference could relate to the origin of radio halos, e.g. details of merger dynamics. Almost radio halos are observed in galaxy clusters which are undergoing sub-cluster merging.

The origin of CR electrons responsible for the radio emission is still under debate. The CR electrons can diffuse out from a point source for a distance of the diffusion velocity $v_d$ times the CR life time $\tau$. The inverse Compton energy loss time of CR electrons which radiate at $\nu \sim$ several $\times 100$~MHz is $\tau \sim 0.1$~Gyr. The diffusion velocity would be the Alfv\'{e}n velocity $\sim 100$~km/s. The diffusion distance is, therefore, $\sim$ 10~kpc, which does not seem as large as the radio halo extent. This problem suggests CR injection and/or acceleration in the ICM. One possible mechanism is the particle acceleration by intracluster magnetic turbulence generated through sub-cluster mergers.

X-ray and radio observations for radio-halo clusters indicate a positive correlation between X-ray luminosity and monochromatic radio power at 1.4~GHz. In addition, the hotter clusters tend to have flatter radio spectra. These correlations suggest the particle acceleration is related with ICM dynamics \citep{2001MNRAS.320..365B,2002ApJ...577..658O,2003ApJ...584..190F}. Sub-cluster mergers would supply energy both to the hot gas and the non-thermal particles through shocks and derived turbulence.

Depolarization and Faraday tomography will be possible through wide-band observations with the SKA. Measurements with depolarization and Faraday tomography can enable us to study three dimensional magnetic field structures in radio-halo clusters. Furthermore, the X-ray satellite Hitomi would reveal ICM turbulence in the veolcity space with its high spectral resolution. These new observations will bring significant progress in understanding the evolution of the magnetic turbulence in the ICM and the particle acceleration during sub-cluster merging.

%%%%%%%%%%%%%%%%%%%%%%%%%%%%%%%%%%%%%%%%%%%%%%%
\subsubsection{Use Case: SKAJP-MAG-7 (CHL)}
\label{mag.s3.ss7.sss2}

\begin{table}
\tbl{Key Requirements of SKAJP-MAG-7 (CHL)}{%
\begin{tabular}{llll}
\hline
Central frequency (Band) & 700 MHz (Band-1) \\
Spectral coverage and resolution & 700 MHz, 10 MHz\\
Angular resolution (baseline) & $15''$ (6.7 km)\\
Image rms (I, P) and bandwidth & (7, 0.7) $\mu$Jy/beam, 10 MHz\\
FoV and pointings & 2.8 deg$^2$, 6 (radio halos)$^\dagger$\\
Total on-source time & 904 hours\\ 
All-sky Band-2 & -- \\
\hline
Central frequency (Band) & 1355 MHz (Band-2) \\
Spectral coverage and resolution & 810 MHz, 100 MHz\\
Angular resolution (baseline) & $15''$ (3.4 km)\\
Image rms (I, P) and bandwidth & (0.9, 0.09) $\mu$Jy/beam, 100 MHz\\
FoV and pointings & 0.75 deg$^2$, 6 (radio halos)$^\dagger$\\
Total on-source time & 855 hours\\
All-sky Band-2 &  -- \\
\hline \end{tabular}}\label{tab:SKAJP-MAG-7}
\begin{tabnote}
$^\dagger$The targets of this project are: 1E0657-56 (the bullet cluster), Abell 209, Abell 520, Abell 521, Abell 1300, and Abell 2744. Integration times are 25, 77, 19, 308, 149, 326 hours (Band-1), and 531, 116, 45, 83, 40, 40 hours (Band-2) shown in this order.
\end{tabnote}
\end{table}

We propose deep observations of radio halos in six galaxy clusters so as to elucidate magnetic turbulence and particle acceleration in the clusters. We focus on the spectral index of the emission, because its spatial distribution correlates with the spatial distribution of the magnetic turbulence power spectrum, based on the fact that the power spectrum gives the acceleration rate and then the acceleration determines the spectral index with competing with energy loss. Therefore, the goal is to obtain spectral index maps of the radio halo emission, and then examine magnetic turbulence and particle acceleration using the maps and theoretical predictions. Polarization properties such as the polarization angle and degree can be also clues of the configuration of magnetic fields. We compare six radio halos to elucidate universality and diversity of magnetic turbulence and particle acceleration in galaxy clusters.

The project can be divided into many scheduling blocks and the standard observing procedure can be adopted. We do imaging with a $\sim 15''$ pixel resolution to resolve structures with the physical scale down to 50 -- 70~kpc, which corresponds to the diffusion distance of relativistic electrons (the Alfv\'{e}n velocity of $\sim 100$~km/s and the energy loss time of $\sim 0.1$~Gyr) for our representative target 1E0657-56 located at $z\sim 0.3$. We are proposing deeper observations than the all-sky Band-2 survey. Data of the all-sky Band-2 survey will be helpful for us to tune up our observing strategy, e.g. we can adjust an observing time for each target.

We realize that some of observations in this use case require the integration time more than a few hundred hours (Table~\ref{tab:SKAJP-MAG-7}). This implies that we must require a high calibration stability throughout a few weeks. We also need to check the confusion limit for such ultra-deep observations, though it might not be critical at the $\sim 15''$ resolution. With SKA2, we will be able to perform the similar observations with much shorter integration time ($\sim$ several hours).

%%%%%%%%%%%%%%%%%%%%%%%%%%%%%%%%%%%%%%%%%%%%%%%
%%%%%%%%%%%%%%%%%%%%%%%%%%%%%%%%%%%%%%%%%%%%%%%
\subsection{High Frequency Emission of 20 Radio Relics}
\label{mag.s3.ss8}

%%%%%%%%%%%%%%%%%%%%%%%%%%%%%%%%%%%%%%%%%%%%%%%
\subsubsection{Objective}
\label{mag.s3.ss8.sss1}

It is believed that CR electrons in radio relics are accelerated at shocks associated with cluster formation \citep{1998A&A...332..395E, 2001ApJ...563..660F}, which is consistent with the facts that significant fractional polarization is often observed and that shock structures are found in X-ray observations \citep{2013PASJ...65...16A}. Diffusive shock acceleration (DSA) \citep{1978ApJ...221L..29B, 1987PhR...154....1B} is the most promising acceleration process, where a radio spectral index is determined from the shock properties. However, the CR electrons lose their energy within a short time and their spectrum changes with the positions behind the shock. Because their characteristic spatial scales can be smaller than the beam size, the data can be degraded because of the low resolution observations. Actually, a spectral index unaffected by the cooling is obtained only for a few relics \citep{2010Sci...330..347V}. In addition, two problems, which might be caused by this effect, are known as follows; (i) Radio and X-ray observations lead to shock properties inconsistent with each other \citep{2015PASJ...67..113I}. (ii) Low Mach number shocks ($M<5$), which are suggested by the radio and X-ray, cannot explain the radio intensity.

We propose wideband and high spatial resolution radio polarimetric observations of relics with the SKA. It is crucial to observe the spectra in the lower frequency bands, which are less influenced by the cooling. Observations in higher frequency bands are also important to know cooling processes and the highest energy of the electrons. Thus, very wideband and high spatial resolution observations are needed, which can be performed only by the SKA.

Synergy with wide FOV X-ray satellites such as eROSITA and DIOS is expected. Wide FOV and large effective area of the SKA will enable us to reveal dependence of the relic statistics on the redshift. Even in the era of SKA pathfinders, two thousands of relics would be found within $z<0.5$ \citep{2012MNRAS.420.2006N}. Combination of the SKA with X-ray will enable us to estimate how energy relevant to cluster mergers are divided into various channels in a cosmological time scale, which is one of the most important keys to understand structure formation. Similar discussion is expected in radio halo studies.

We can study 3D structures of the fields with Faraday tomography technique for high spatial resolution data little affected by the beam depolarization. High fractional polarization (50 -- 60~\%) is observed in some relics. Considering the foreground magnetic field, their intrinsic value is likely even higher. This suggests the ordered magnetic fields in a Mpc scale.  How such structures form might be a clue to understand the evolution of the magnetic field in clusters and large scale structures. It is expected we can remove the foreground magnetic fields effects with depolarization analysis. In addition, combined with 3D structures revealed with the tomography, we can study their origin and properties in more detail.

%%%%%%%%%%%%%%%%%%%%%%%%%%%%%%%%%%%%%%%%%%%%%%%
\subsubsection{Use Case: SKAJP-MAG-8 (CLC)}
\label{mag.s3.ss8.sss2}

\begin{table}
\tbl{Key Requirements of SKAJP-MAG-8 (CLC)}{%
\begin{tabular}{llll}
\hline
Central frequency (Band) & 9200 MHz (Band-5) \\
Spectral coverage and resolution & 9200 MHz, 10 MHz \\
Angular resolution (baseline) & $16.''2$ (0.47 km)\\
Image rms (I, P) and bandwidth & (5.7, 5.7) $\mu$Jy/beam, 10 MHz\\
FoV and pointings & 0.016 deg$^2$, 2 $\times$ 20$^\dagger$ \\
Total on-source time & 220 hours\\
All-sky Band-2 & Essential$^\ddagger$ \\
\hline
\end{tabular}}\label{tab:SKAJP-MAG-8}
\begin{tabnote}
$^\dagger$ Each of 20 radio relics is observed with two pointings (Mosaicking).\\
$^\ddagger$ Use to select sources and check depolarization.
\end{tabnote}
\end{table}

We propose observations of 20 radio relics with SKA1-MID Band-5. A radio spectral shape in a high frequency part provides us with useful information about a high energy part of the cosmic-ray electron spectrum, which can constrain the particle acceleration processes and magnetic fields in the radio relic. We search for a break or cut-off of the radio spectra in a higher frequency part. Polarization properties such as the polarization angle and polarization degree can be used to find the configuration of magnetic fields. We compare the high frequency feature of 20 radio relics to elucidate universality and diversity of acceleration mechanisms in radio relics.

The standard observing procedure can be adopted. The project can be divided into 40 scheduling blocks for the 20 targets with appropriate mosaicking strategy, where each target is observed with two pointings. For instance, our representative target (toothbrush radio relic, z=0.225 \cite{2012A&A...546A.124V}) extends over $2' \times 8'$, while one side of the FoV is $6.6'$ at Band-5. Note that antenna baselines of $\sim 30$~m would provide the maximum angular scale of $3'.6$ at 9200~MHz \citep{2014skao.rept.....G}. Data of the all-sky Band-2 survey is essential; we suppose that we have radio-relic candidates in the survey catalog and we select 20 best targets from the candidates.

%%%%%%%%%%%%%%%%%%%%%%%%%%%%%%%%%%%%%%%%%%%%%%%
%%%%%%%%%%%%%%%%%%%%%%%%%%%%%%%%%%%%%%%%%%%%%%%
\subsection{Magnetic Turbulence in 42 Nearby Galaxy Clusters}
\label{mag.s3.ss9}

%%%%%%%%%%%%%%%%%%%%%%%%%%%%%%%%%%%%%%%%%%%%%%%
\subsubsection{Objective}
\label{mag.s3.ss9.sss1}

Three dimensional structures of cluster magnetic fields can be complicated due to turbulence. It is nontrivial to reconstruct magnetic turbulence from projected observables such as synchrotron emission and Faraday rotation. Comparison between observed data and theoretical models is helpful for resolving the complexity. The comparison also makes quantification of magnetic fields easy. While simple models of uniform, random magnetic fields have been considered in the literature, more sophisticated models were proposed recently \citep{2003A&A...401..835E,2004A&A...424..429M}. The models assume random gaussian fields and the spatial power spectrum with cutoffs at the minimum and maximum scales. They also assume a correlation between the gas density and magnetic-field strength. Such models allow to quantify magnetic fields in detail from statistics of observed RMs; Kolmogorov-like turbulence was suggested to be present \citep{2008A&A...483..699G}.

Although studies for individual clusters will advance our understanding of the nature of magnetic fields, it would be insufficient to elucidate the evolution history of them. More strategic investigation is required to address the time domain, and observational tests of theoretical hypotheses could become a powerful probe. Here, whatever the origin is, cluster magnetic fields should be affected by turbulence induced by cluster mergers, motions of cluster galaxies, and AGN. Cluster magnetic fields should be also altered, when radiative-cooling inflows of the gas take place in the cluster center. Therefore, we suspect that magnetic turbulence correlates with cluster morphology.

Motivated by the dynamo theory (e.g., \cite{2008Sci...320..909R}), we construct a theoretical hypothesis and test it with SKA observations. Our hypothesis is that X-ray morphology indicates the ``age" of the cluster. Up to now, galaxy clusters have been classified into irregular, regular, and cool-core types according to the characteristics of X-ray surface brightness and temperature. We regard these types as features in different evolution stages: merger, relaxation, and cooling, in this order. Each stage may correspond to amplification, saturation, and decay of turbulence in the context of dynamo action. If this is the case, clusters with different X-ray morphologies should represent different properties of magnetic fields, and we can statistically examine it once we have a sufficient number of samples. So far, there are over 100 clusters whose X-ray morphology are known \citep{1999ApJ...517..627M, 2002ApJ...567..L23O,2005PASJ...57..419A}. Radio counterparts are, however, very limited. We need SKA's amazing sensitivity to study all of these clusters. If our hypothesis is proven by SKA observations, it provide us important suggestions of co-evolution of the ICM with magnetic fields and origin of the cluster magnetic fields. 

\begin{figure}[tbp]
\begin{center}
\FigureFile(80mm,80mm){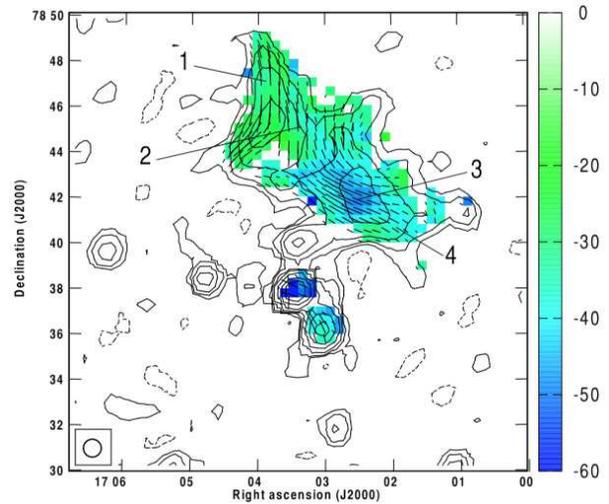}
\end{center}
\caption{
RM distribution of the Abell 2256 radio relics with the intrinsic B-vector as black bars. Black contours show the total intensity at 2051 MHz (see \cite{2015PASJ...67..110O} for details).
}
\label{f11}
\end{figure}

Our focus on magnetic turbulence matches quite well with depolarization and Faraday tomography, because they can be sensitive to the expected RM dispersion of several tens rad~m$^{-2}$ at the SKA observing frequency (Fig. \ref{f01}). We have already carried out JVLA pilot observations of six galaxy clusters with different X-ray morphologies. Although our analysis is still in progress, our first release for Abell 2256 (\cite{2015PASJ...67..110O}, Fig.~\ref{f11}) successfully demonstrated that depolarization is useful to examine turbulent magnetic fields and Faraday tomography has the potential to resolve multiple components along the LOS.

%%%%%%%%%%%%%%%%%%%%%%%%%%%%%%%%%%%%%%%%%%%%%%%
\subsubsection{Use Case: SKAJP-MAG-9 (CTB)}
\label{mag.s3.ss9.sss2}

\begin{table}
\tbl{Key Requirements of SKAJP-MAG-9 (CTB)}{%
\begin{tabular}{llll}
\hline
Central frequency (Band) & 1355 MHz (Band-2) \\
Spectral coverage and resolution & 770 MHz, 10 MHz \\
Angular resolution (baseline) & $1''$ (46.2 km)\\
Image rms (I, P) and bandwidth & (4, 4) $\mu$Jy/beam, 10 MHz\\
FoV and pointings & 0.75 deg$^2$, 42 nearby clusters \\
Total on-source time & 114 hours\\
All-sky Band-2 & Desirable$^\dagger$ \\
\hline
\end{tabular}}\label{tab:SKAJP-MAG-9}
\begin{tabnote}
$^\dagger$ Use to check the existence of polarization in the targets.
\end{tabnote}
\end{table}

We propose observations of the 42 nearby galaxy clusters whose X-ray morphologies are known and their declinations are less than the limit, $+44^\circ$. We systematically explore synchrotron emission and polarization from these clusters as well as polarized sources embedded in/behind these clusters. We examine polarization properties such as the polarization angle, the fractional polarization, and RM. Based on X-ray morphologies such as irregular and cool-core structures, we classify the 42 clusters into some categories (irregular, regular, cool-core, Bautz-Morgan type, etc) and compare radio properties among the categories. From the comparisons, we argue the evolution of magnetic turbulence. SKA1-MID Band-2 provides the best specification; a high resolution ($1'' \sim 1$~kpc in average), a sufficient image size ($49' \sim 3$~Mpc one side), and a satisfactory maximum angular scale ($20' \sim~1.2$~Mpc at 1355~MHz for antenna baselines of $\sim 30$~m). 

The standard observing procedure can be adopted. Observations of the clusters toward which no radio emission is reported from the all-sky Band-2 survey may be cancelled. In order to identify the contribution to the RM from the foreground and background, Faraday tomography with multi frequency observations would be very promising. It is also important to interpret the Faraday spectra which are calculated from the observed and theoretical data to understand the evolution of the turbulent magnetic fields. Therefore, with I, Q, and U, Faraday synthesis is carried out followed by Faraday RM clean with standard parameters. In addition, for each source, Q, and U are fitted with several FDF models by means of an MCMC method, where the result of RM clean is used as an initial guess. Here, a 10 MHz channel provides a maximum Faraday depth of about 4350 rad~m$^{-2}$ at 1355 MHz.

%%%%%%%%%%%%%%%%%%%%%%%%%%%%%%%%%%%%%%%%%%%%%%%
%%%%%%%%%%%%%%%%%%%%%%%%%%%%%%%%%%%%%%%%%%%%%%%
\subsection{Extraction of RM in the LSS}
\label{mag.s3.ss10}

%%%%%%%%%%%%%%%%%%%%%%%%%%%%%%%%%%%%%%%%%%%%%%%
\subsubsection{Objective}
\label{mag.s3.ss10.sss1}

\begin{figure}[tbp]
\begin{center}
\FigureFile(75mm,75mm){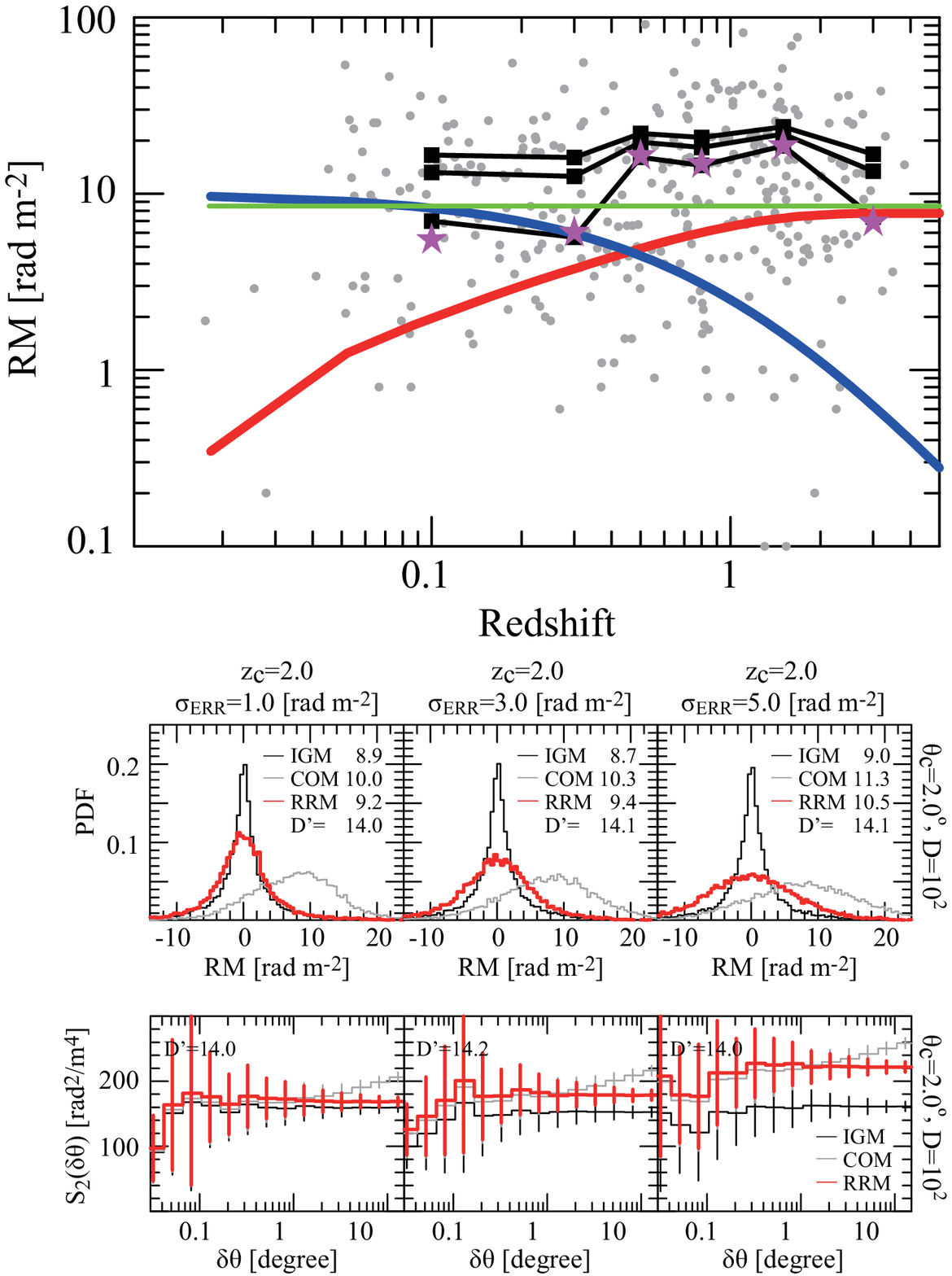}
\end{center}
\caption{
(Top) Redshift dependence of RM \citep{2014ApJ...790..123A}. The gray points are 317 extragalactic sources at $|b|> 75^\circ$ \citep{1209.1438v3}. The black lines with marks are standard deviations of RM for the shown redshift subsets, where the random noise of 0, 10, and 15 rad~m$^{-2}$ were subtracted. The blue, red, and green lines are theoretical expectations of the intrinsic RM associated with a extragalactic polarized source, RM due to the IGMF \citep{2011ApJ...738..134A}, and RM due to the Milky Way \citep{2010MNRAS.409L..99S}, respectively. The purple marks are expectations of RM due to DIGs, where the random observational noise of 15 rad~m$^{-2}$ is assumed. (Bottom) A simulations of South galactic pole observation; the probability distribution function of RM (upper) and the second order structure function of RM (lower). The random noise with the RM standard deviation of 1, 3, 5 rad~m$^{-2}$ are considered from the left to right figures, respectively. See \citep{2014ApJ...790..123A} for details.
}
\label{f12}
\end{figure}

We focus on RM due to the IGMF in the large-scale structure (LSS) rather than synchrotron emission from the LSS, based on previous works by Japanese researchers. As already described in \S\ref{mag.s2.ss7}, the IGMF would be very weak (see also \S\ref{mag.s3.ss11}). Hence it is not easy to estimate only the IGMF in a single LOS, since individual contributions of external polarized source, our galaxy, and so on, to RM are not negligible in general. It is valid to adopt statistical approaches to remove such contributions. Moreover, the statistics such as average, variance and second-order structure function of RMs would have some characteristics of RM due to the IGMF. The Japan's advantage for the study is that we have a theoretical model of the IGMF and abundant knowledge about foreground and background components introduced in the previous sections. Our goal is to extract the characteristics after choosing ideal polarized sources from an enormous number of background sources, by means of the methods proposed below. 

We have already simulated the estimation of the IGMF, where model RM maps are constructed following previous works. \citet{2014ApJ...790..123A} randomly allocated polarized sources on the plane of the sky accordance with the redshift distribution of observed polarized sources, and chose half of them at random as the one without intervening galaxies based on the observed statistics. Then, they further removed the sources with $z<2$, for which the intrinsic RMs can be significant, as well as the one including galaxy clusters along the LOS. After that, they chose the sources with small observational errors. Finally, they applied a high-pass filter to remove Galactic foreground RM. As a result, it is found that the variance of RM of the IGMF can be estimated with sufficient accuracy with the statistics obtained by POSSUM, the polarization survey planned with ASKAP (100 polarized sources per 1 square degree with the sensitivity of a few $\mu$Jy/beam, Fig.~\ref{f12}). The POSSUM would be, however, hard to reveal the structure function of the IGMF (Fig.~\ref{f12}). We require the SKA's dense RM grids to find a typical scale of the structure of the IGMF.

\begin{figure}[tbp]
\begin{center}
\FigureFile(80mm,80mm){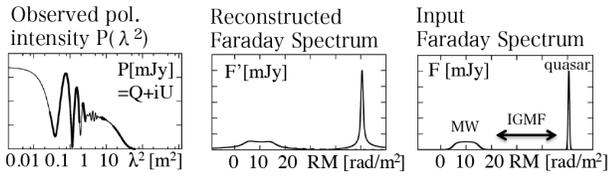}
\end{center}
\caption{
Exploring the IGMF by means of Faraday Tomography. (left) the mock polarized intensity spectrum, (middle) the reconstructed Faraday spectrum with the SKA2, (right) the input Faraday spectrum. see \citet{2014PASJ...66...65A} for details.
}
\label{f13}
\end{figure}

Faraday tomography has the potential to separate foreground and background along a LOS (e.g., \cite{2015ApJ...815...49A}) and extract RM of only the IGMF. However, as introduced in the previous section, it is necessary to understand the best situation of observation, and to interpret the reconstructed Faraday spectra. In other words, understanding the nature of depolarization and Faraday tomography on this specific subject is essential. \citet{2014PASJ...66...65A} discussed the possible methodology for estimating RM of the IGMF using Faraday tomography. Their idea is to observe a distant polarized source through the Galaxy and to measure RM of the IGMF as the gap of FDF (Fig.~\ref{f13}). They simulated some situations with a simple model. They found that, if the RM of the IGMF is larger than $\sim 10$ rad~m$^{-2}$, the gap will be detectable directly from the FDF within a factor of 2 in the SKA era. Moreover, \citet{2014ApJ...792...51I} found that a model fitting method called QU-fit (as it execute fitting using stoke Q and U) is capable to estimate the IGMF with SKA precursors such as ASKAP even if the RM is $\sim 5$ rad~m$^{-2}$.

The key requirements to perform the methods described above is (i) to obtain an enormous number of background polarized sources that allow us to find ideal sources, and (ii) to obtain wideband polarization spectra that allows us to further examine ideal sources by means of depolarization and Faraday tomography. In order to clear these requirements, we need to carry out wideband all-sky polarization surveys.

%%%%%%%%%%%%%%%%%%%%%%%%%%%%%%%%%%%%%%%%%%%%%%%
\subsubsection{Use Case: SKAJP-MAG-10 (LSS)}
\label{mag.s3.ss10.sss2}

\begin{table}
\tbl{Key Requirements of SKAJP-MAG-10 (LSS)}{%
\begin{tabular}{llll}
\hline
Central frequency (Band) & 700 MHz (Band-1) \\
Spectral coverage and resolution & 770 MHz, 1 MHz \\
Angular resolution (baseline) & $2''$ (45 km)\\
Image rms (I, P) and bandwidth & (16, 16) $\mu$Jy/beam, 210 MHz\\
FoV and pointings & 2.8 deg$^2$, 12903$^\dagger$ Mosaicking\\
Total on-source time & 418 hours\\
All-sky Band-2 & Essential$^\dagger$ \\
\hline
\end{tabular}}\label{tab:SKAJP-MAG-10}
\begin{tabnote}
$^\dagger$ This is a follow-up all-sky survey.
\end{tabnote}
\end{table}

We propose the Band-1 polarization survey as a follow-up of the Band-2 survey. As mentioned above, one of our strategy for finding the IGMF is to observe a distant polarized source through the Galaxy. The Faraday thin source is suitable for finding RM of the IGMF as the gap of FDF. Since lower frequency observation has better sensitivity for the Faraday thin sources, the Band-1 survey is essential to prove the IGMF. The polarization information with broader frequency coverage also allows us to obtain the sensitivity of Faraday structure as well as more precise RM estimation if the situation has the simple Faraday structure.

In concert with the Band-2 survey, the beam size is selected to $2''\times 2''$. For the image rms level, we require 16~$\mu$Jy/beam with 210 MHz bandwidth based on the Band-2 survey with taken into account the typical radio galaxies ($\alpha\sim -1$). We finally request 418 hours on-source observing time. The project can be divided into many scheduling blocks and the standard observing procedure can be adopted. The project could be commensal with HI, cosmology, and continuum survey. In this case, we need to consult with these survey teams about observing strategy.

We will perform the standard polarization analysis. Here, once a source is detected, the bandwidth is recursively divided by half down to the 1 MHz channel or down to the detection limit so as to gain spectroscopic data. Here, a 1 MHz channel provides a maximum Faraday depth of about 5000 rad~m$^{-2}$ at 700 MHz. For a given Stokes I, Q, and U, the estimation of RM and Faraday tomography are performed. The correlation of the LSS with RM and Faraday structure is investigated. In addition, for the interested situation where there seems to be the IGMF, the more sophisticated method such as the QU-fitting with the replica exchange MCMC algorithm is applied for the more detailed study.

%%%%%%%%%%%%%%%%%%%%%%%%%%%%%%%%%%%%%%%%%%%%%%%
%%%%%%%%%%%%%%%%%%%%%%%%%%%%%%%%%%%%%%%%%%%%%%%
\subsection{Cosmological Magnetic Fields}
\label{mag.s3.ss11}

%%%%%%%%%%%%%%%%%%%%%%%%%%%%%%%%%%%%%%%%%%%%%%%
\subsubsection{Objective}
\label{mag.s3.ss11.sss1}

Our ultimate question of cosmic magnetism would be ``What is the seed of Universe's magnetic fields?". Although the question is yet far beyond the scopes of any observational projects, theorists have addressed this frontier. We summarize some of major possible origins of Universe's magnetic fields below (see \cite{2012SSRv..166....1R,2012SSRv..166...37W} for reviews).

\begin{itemize}

\item{\bf Inflation Era:}
Since electromagnetic field is conformal transformation invariant, there is the principle that electromagnetic quantum fluctuation cannot be generated in the standard inflation model. Hence, beyond the standard model, magnetic-field generation in the inflation era has been argued together with the additional scalar field such as Dilaton \citep{2004PhRvD..69d3507B} and Higgs \citep{2004PhRvD..70d3004P}, as well as the field of gravity \citep{1988PhRvD..37.2743T}. They can generate largest-scale coherent magnetic fields but with tiny amplitude of $10^{-15}$ -- $10^{-25}$~G at the present epoch. There are significant contributions from Japanese researches \citep{2009JCAP...12..009K, 2012PhRvD..86b3512S,2014JCAP...03..013F}, in particular, to the back-reaction problem. In short, field generation must be suppressed at largest scales; $B\lesssim 10^{-30}$~G at Mpc for the simplest $f^2(\phi)F_{\mu\nu}F^{\mu\nu}$ type models \citep{2014JCAP...06..053F}. 

\item{\bf Reheating and phase transition Era:}
There is, on the other hand, no problem on suppression of largest magnetic fields in the field generation at cosmic reheating \citep{2014JCAP...05..040K} and/or phase transition \citep{1983PhRvL..51.1488H,1991PhLB..265..258V}. \citet{2013PhRvD..87h3007K} argued that the field with strength and coherence length of $10^{-9}$~G and $\sim$50 kpc can be generated at QCD phase transition, and $10^{-10}$~G and $\sim$0.3 kpc at QED phase transition, respectively. These generate significant power at small scales, compared to those in the inflation era.

\item{\bf Recombination Era:}
Magnetic fields can be generated at cosmological recombination \citep{2005PhRvL..95l1301T,2006Sci...311..827I}. In the primordial plasma, frequent Thomson scattering can separate electrons from protons and lead generation of electric fields for the protons to catch up with the electrons. The induced electric fields can then generate magnetic fields according to the Maxwell equations. 

\item{\bf First-Star and EoR Era:}
Likewise, the electric field can be also induced by UV radiation from the first-generation stars as well as by CRs produced by supernova explosion of the first-generation stars \citep{2000ApJ...539..505G, 2005A&A...443..367L, 2010ApJ...716.1566A, 2011ApJ...729...73M}. Such seed fields generated before the large-scale structure formation would evolve to $O(1$ -- $100)$~nG magnetic fields in filaments \citep{2008A&A...482L..13D}.

\item{\bf Galaxy-Evolution Era:}
Undoubtable sources of magnetic fields are instabilities and dynamo actions in galaxies (e.g., \cite{2013ApJ...764...81M}). It is likely that magnetic fields and CRs escape from galaxies into the intergalactic space through jets, supernovae, winds, and ram pressure stripping. Cosmological simulations of galaxy formation suggest that the leakage result in $B\sim 0.1$ -- 10~$\mu$G in galaxy clusters \citep{2009MNRAS.392.1008D, 2009ApJ...698L..14X, 2010MNRAS.408..684S}.

\item{\bf Structure-Formation Era:}
Cosmological shock waves in the structure formation can generate seed magnetic fields by the Biermann battery \citep{1998A&A...335...19R}, the Wibel instability \citep{2003ApJ...599..964O}, and plasma instability \citep{2005MNRAS.364..247F}. Seed magnetic fields of any origins could be further amplified through compression and turbulence dynamo. A conservative range of the resultant IGMF in filaments would be $O(1$ -- $100)$~nG \citep{2008Sci...320..909R}, with the coherence length $O(100)$~kpc \citep{2009ApJ...705L..90C}.

\end{itemize}

Some of the SKA-JP Magnetism SWG members have investigated generation of magnetic fields at cosmological recombination \citep{2005PhRvL..95l1301T,2006Sci...311..827I}. Here, the key point to note is that the electric fields should have rotation component to generate magnetic fields \citep{1970MNRAS.147..279H}. Therefore the standard density perturbations can not be responsible for the generation of magnetic fields at first order. For a possible generation of magnetic fields based on the standard cosmological perturbation theory, we need to go beyond the first order or consider an external source such as textures and cosmic strings to generate vector mode perturbations. Because the cosmological perturbation theory is well established, we can make a firm prediction about the amplitude and the spectrum of the generated magnetic fields.

\begin{figure}[tbp]
\begin{center}
\FigureFile(80mm,80mm){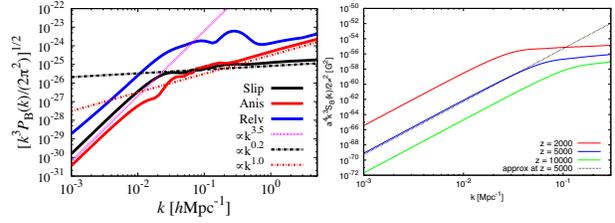}
\end{center}
\caption{
The spatial power spectra of magnetic fields generated at the epoch of reionization \citep{2015PhRvD..91l3510S,2016arXiv160101059H}. The left and right panels show the cases of primordial density perturbation and texture topological defect, respectively.
}
\label{f14}
\end{figure}

In Fig.~\ref{f14} we show some examples of magnetic field spectra at recombination. The amplitude of generated magnetic fields from second order density perturbations can reach at most $B\approx 10^{-23}$~Gauss on $k=0.5h$~Mpc$^{-1}$ scale at $z=500$ \citep{2015PhRvD..91l3510S}, while the magnetic fields from texture of cosmic string can be as large as $10^{-20}$~Gauss \citep{2015JCAP...04..007H}. We find that the amplitude is too small to directly explain the magnetic fields observed today, but these fields might be a seed for the subsequent dynamo amplification.

%%%%%%%%%%%%%%%%%%%%%%%%%%%%%%%%%%%%%%%%%%%%%%%
%%%%%%%%%%%%%%%%%%%%%%%%%%%%%%%%%%%%%%%%%%%%%%%
\subsection{Advancing Faraday Tomography}
\label{mag.s3.ss12}

%%%%%%%%%%%%%%%%%%%%%%%%%%%%%%%%%%%%%%%%%%%%%%%
\subsubsection{Objective}
\label{mag.s3.ss12.sss1}

Faraday tomography is a powerful way to probe 3D structure of galaxies and intergalactic magnetic fields \citep{2014PASJ...66...65A,2014PASJ...66....5I}. However, this method needs a broadband polarimetry and has been studied only recently. We are doing studies related to both the practical application and fundamental properties of Faraday tomography.

We are developing a software of Faraday tomography and it has already been incorporated into the pipeline of POSSUM, which is an all-sky polarization survey of ASKAP. The software is based on QU-fitting where the Faraday spectrum is assumed to be a combination of Gaussian functions and is fitted to the observed polarization spectrum. The fitting is performed by Markov Chain Monte Carlo method and, as a result of the fitting, the model parameters such as the Faraday depths, amplitudes, polarization angle and widths of the Gaussian components are obtained. Using information criteria such as AIC and BIC, the number of Gaussian components is determined.

In the SKA era, the number of polarized objects to be analyzed is enormous so that the computational cost must be reduced as much as possible. Furthermore, the conventional information criteria currently used does not work well in cases with the presence of multiple sources located closely in the Faraday depth space \citep{2015AJ....149...60S,Nakagawa2016}. To solve these problems, we are planning to improve our software with adopting a more sophisticated algorithm such as replica exchange method.

\begin{figure}[tbp]
\begin{center}
\FigureFile(75mm,75mm){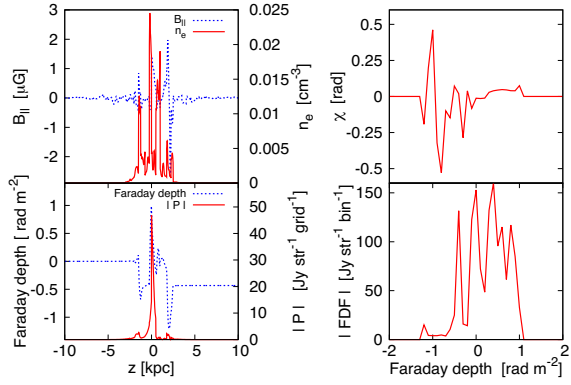}
\end{center}
\caption{
The Faraday spectrum of a simulated face-on spiral galaxy \citep{2014ApJ...792...51I}. (Top left) distributions of the LOS component of magnetic fields and the thermal electrons, (bottom left) distributions of the polarized intensity and Faraday depth, (top right) the polarization angle, (bottom right) the absolute Faraday spectrum.
}
\label{f15}
\end{figure}

In utilizing QU-fitting, the functional form of the model of Faraday spectrum is a critical assumption. The realistic form of the Faraday spectrum was studied by calculating it from galactic models (\cite{2014ApJ...792...51I}, Fig.~\ref{f15}) and it was shown that the Faraday spectrum of Galaxies can be very complicated and cannot be approximated by Gaussian or other simple analytic function. The complexity is mainly due to the stochasticity of turbulence in magnetic fields and gas density. \cite{Tashiro2015} showed that the stochasticity can be significantly reduced by using a large beam and the global properties of galaxies such as coherent magnetic fields and characteristic scale of turbulence could be extracted from the Faraday spectrum.

Finally, RM CLEAN is another standard method of Faraday tomography. This method does not need any assumption on the functional form of Faraday spectrum. However, due to limited frequency coverage, it is difficult to reconstruct Faraday spectrum especially when multiple polarized sources are located closely in Faraday depth space \citep{2014PASJ...66...61K}. In \cite{Miyashita2015}, we proposed the chi-square value calculated from the observed polarization spectrum and the result of RM CLEAN as a criterion for judging if RM CLEAN is working well or not.

%%%%%%%%%%%%%%%%%%%%%%%%%%%%%%%%%%%%%%%%%%%%%%%
%%%%%%%%%%%%%%%%%%%%%%%%%%%%%%%%%%%%%%%%%%%%%%%
%%%%%%%%%%%%%%%%%%%%%%%%%%%%%%%%%%%%%%%%%%%%%%%
\section{Summary}
\label{mag.s4}

The origin and evolution of cosmic magnetism is one of the fundamental questions in modern astrophysics. Centimeter and meter radio observation is an established method to study magnetic fields. We, SKA-JP Magnetism SWG, consider that it is important to investigate magnetic fields in various astronomical objects. We expect that depolarization and Faraday tomography, which will become applicable in the SKA era, allow us to resolve magnetic fields in the four dimensional space, improving our understanding of magnetic fields dramatically. We suggest that the SKA should possess supreme sensitivity and wide frequency coverage to enhance the study of depolarization and Faraday tomography. More specifically, we request the following ``core" specifications of the SKA. 

\begin{itemize}

\item {\bf Continuous bandwidth of at least 0.3 -- 3 GHz:} The frequency coverage is an essential specification for depolarization and Faraday tomography. Although it can be achieved by performing multiple observations with narrow simultaneous bandwidths, it is much preferable to deploy a few wideband single pixel feeds (WBSPFs) or several phased array feeds (PAFs) so as to dramatically improve the efficiency of the survey. It would be also effective to adopt a sampler which is possible to sample data unevenly in frequency and evenly in frequency-squared. 

\item {\bf All-sky 1 $\mu$Jy survey + 400 deg$^2$ 0.1 $\mu$Jy deep fields:} Our use cases can be summarized into two big survey projects - all-sky survey and deep survey toward high galactic latitudes. With SKA2-MID (0.5 deg$^2$ FoV), all-sky 1 $\mu$Jy sensitivity survey can be achieved in $\sim 1000$ hours. With a similar observing time, we can perform deep fields of 0.1 $\mu$Jy for in total 400 deg$^2$ FoV toward North and South Galactic poles. Such a long-time survey can only be allowed as a KSP under the current operation concept of the SKA. In order to carry out any KSP as a principle investigator, we have to join the SKA project as a full member.

\item {\bf Angular resolution of 0.1 arcsecond at 1 GHz:} A high angular resolution can relax beam depolarization and increase the possibility to detect more polarized sources. Such an improvement is beneficial on the studies of MHD turbulence in the Milky Way and galaxy clusters. The SKA1 will provide about $1''$ resolution at 1 GHz by the maximum baseline of $\sim 100$ km. In order to provide distinguished progress from SKA1 to SKA2, the SKA2 should achieve $0.''1$ resolution at 1 GHz, corresponding maximum baseline of $\sim 1,000$ km.

\item {\bf Largest angular scale of 1 degree at 3 GHz:} With the minimum baseline 20 meter at zenith, the largest angular scale is around $30'$ at 1.5 GHz. This limit may not be a serious matter for many of sciences we have proposed. However, some sciences addressing diffuse emission may require to detect far larger-scale structures. In order to remove an uncertainty due to the missing flux, we request the largest angular scale of $1^\circ$ at 3 GHz. This requirement can be achieved by deploying dishes with the minimum baseline of a few meter or deploying a very large single dish. This is beyond the current vision of the SKA project and could be a Japanese unique proposal.

\item {\bf Robust and fast science data processor pipeline:} It is impossible to look around a few hundred million sources manually, so that development of robust and fast science data processor (SDP) pipelines is crucial for the SKA project. We need to understand how we can make the SDP pipeline robust, fast, and automated. Because both theory and application of Faraday tomography are still less established, we should be responsible for the development in order that we can achieve our science cases with Faraday tomography. It may require substantial funding and human resources.

\end{itemize}

Finally, we could obtain much more science outcomes from the following synergies with other large projects. 

\begin{itemize}

\item{\bf HI observations:} There is a long history of studying molecular clouds with Nobeyama 45 m and ALMA. It was suggested in M33 that magnetic fields of molecular clouds are aligned with global magnetic fields \citep{2011Natur.479..499L}. On the other hand, the star formation rate is not correlated to the strength of regular magnetic fields but to the strength of random magnetic fields \citep{2013A&A...557A.129T}. The SKA commensal observations of HI and polarization would be important to clarify the impact of global magnetic fields on small-scall star formation.

\item{\bf EoR 21 cm line observations:} Synchrotron radiation of the Milky Way is fatal foreground for examining redshifted HI lines at the Epoch of Reionization (EoR). In meter wavelength, time variability of Stokes Q and U is huge and hence calibration of leakages among Stokes I, Q, and U is non-trivial. Making the Milky Way synchrotron template would be useful for the studies of both EoR and cosmic magnetism.

\item{\bf Polarization from the Sun and stars:} Magnetic fields of the Sun has been studied from observations of circular polarization with Nobeyama centimeter heliograph \citep{2013PASJ...65S..14I}. Since the observed characteristic frequency of the polarization depends on the altitude of Solar atmosphere, wideband imaging allows us to study three dimensional structures of magnetic fields and their time variabilities, and we can clarify relations between magnetic fields and observed electromagnetic phenomena. With SKA's high resolution and high sensitivity, we could extend such studies into stars.

\item{\bf Other frequency observations:} ALMA will explore magnetic fields of individual objects by measuring dust polarization and Zeeman effects, while the SKA will provide complementary information of its surrounding. All-sky maps of cosmic microwave background (CMB) experiences and the SKA will be complementary to each other in terms of frequency, and such complementarity can be extended further to the visible band such as SGMAP, a Japanese optical polarimetric survey project. CMB could be even utilized as a source of polarization for the study of magnetic fields \citep{2003ApJ...584..599O}. X-ray experiences such as Hitomi and DIOS will provide supportive data to SKA observations for studies of shocks, turbulence, and CRs in supernova remnants and galaxy clusters.

\item{\bf Ultra high energy CRs:} Telescope Array (TA) Experience, which Japan takes part in, recently found that 72 events with the energy larger than 5.7$\times 10^{19}$ eV in the last 5 years are localized on a specific sky area (called hot spot) at 3.4$\sigma$ (99.963 \%) significance \citep{2014ApJ...790L..21A}. In addition, CRs above $10^{18.2}$ eV are mostly protons. The anisotropy implies deflection of CRs due to magnetic fields in the Milky way and the large-scale structure within up to 250 Mpc. Faraday tomography with the SKA would be quite helpful to understand the magnetic fields and would provide a clue for the origin of ultra-high energy CRs.

\end{itemize}

\vskip 12pt

The authors would like to thank Dr. Hiroki Akamatsu, Dr. Osamu Kameya, Dr. Kohei Kumazaki, Dr. Hajime Susa, and Dr. Dai Yamazaki for their contributions onto Japanese magnetism sciences. Special thanks are as well Dr. Kazumasa Iwai and Dr. Eiji Kido for their contributions onto synergy. The authors are indebted very much to Dr. Makoto Inoue, Dr. Susumu Inoue, Dr. Sachiko Onodera, Dr. Kazuhiro Nakazawa, Dr. Shinpei Shibata, and Dr. Yoshiaki Sofue for useful comments, suggestions, and encouragements. The authors wish to acknowledge Mr. Kenta Anraku, Ms. Ikumi Takahashi, Mr. Yuichi Tashiro, and Mr. Michihiro Takahashi for their helps. The authors are grateful to International SKA Cosmic Magnetism SWG members for providing us opportunities of open discussion and cooperation. This work was supported in part by JSPS KAKENHI Grants: 15K17614 (TA), 15H03639 (TA), 15K05080 (YF), 26400218 (MT). SI was supported by the National Research Foundation of Korea through grant 2007-0093860.

\bibliographystyle{aa}
\bibliography{paper}

\end{document}